\documentclass[a4paper,aps,prd,onecolumn,preprintnumbers,showpacs,nofootinbib]{revtex4}
\usepackage{amsmath,amssymb,graphics,epsfig,subfigure}
\usepackage{color}
\usepackage{float}
\usepackage{hyperref}
\hypersetup{colorlinks=true,linkcolor=blue,citecolor=green}

\begin{document}
\newcommand {\nn} {\nonumber}
\renewcommand{\baselinestretch}{1.3}
\newcommand{\dd}{{\rm{d}}}
\newcommand{\rovno}{\!\!\!\!& = &\!\!\!\!}
\newcommand{\im}{\mathrm{i}}
\newcommand{\Ree}{\mathrm{Re\,}}
\newcommand{\Imm}{\mathrm{Im\,}}

\title{Magnetic field effects on spherical orbit in Kerr-Bertotti-Robinson spacetime: constraints from jet precession of M87* }

\author{Chao-Hui Wang$^{a,b}$}
%\email{wangchh2023@lzu.edu.cn}
\author{Xiang-Cheng Meng$^{a,b}$}
%\email{mxiangcheng2023@lzu.edu.cn}
\author{Shao-Wen Wei$^{a,b}$}
\email{weishw@lzu.edu.cn, corresponding author}
\affiliation{
$^{a}$Key Laboratory of Theoretical Physics of Gansu Province, Key Laboratory of Quantum Theory and Applications of MoE, Gansu Provincial Research Center for Basic Disciplines of Quantum Physics,
Lanzhou University, Lanzhou 730000, People's Republic of China.\\
$^{b}$Institute of Theoretical Physics $\&$ Research Center of Gravitation, Lanzhou Center for Theoretical Physics, School of Physical Science and Technology, Lanzhou University, Lanzhou 730000, People's Republic of China.
}

\begin{abstract}
The recently reported precession period of about $11.24$ years of the M87* jet provides a sensitive probe of strong field gravity and the electromagnetic environment in the immediate vicinity of supermassive black holes. In this work, we study the precession of the spherical orbit in the Kerr-Bertotti-Robinson geometry describing a rotating black hole immersed in a uniform electromagnetic field. Although the timelike geodesics is non-separable, we develop a Hamiltonian approach to investigate the spherical orbits.  For sufficiently strong magnetic fields, the study shows that the spherical orbits can only exist within a finite radial range for given orbital inclination. Requiring the existence of the spherical orbits, we obtain an upper bound of the magnetic field, i.e., $B<0.33  M^{-1}$ for prograde and $B<0.0165  M^{-1}$ for retrograde motion. Furthermore, imposing the observed jet precession period, we obtain a significantly tighter constraint, $B\lesssim 0.0145  M^{-1}$, providing a new constrain on the magnetic field of M87* independent of the shadow. Our results provide unified constraints on the parameters of the KBR black hole and demonstrate that the jet precession offers a robust and complementary probe of magnetized black holes in the strong gravity regime.
\end{abstract}



\maketitle
%\tableofcontents

\section{Introduction}\label{secIntroduction}
Recent unprecedented progress in exploring the strong gravity regime has been driven by two major observational breakthroughs: the direct detection of gravitational waves from binary black hole mergers~\cite{LIGOScientific:2016aoc, LIGOScientific:2016sjg} and horizon scale imaging of supermassive black holes, most notably M87*~\cite{EventHorizonTelescope:2019dse} and Sagittarius~A*~\cite{EventHorizonTelescope:2022wkp}. These observations provide complementary and independent probes of spacetime dynamics in the nonlinear regime of general relativity, opening new avenues for inferring black hole properties from realistic astrophysical data~\cite{EventHorizonTelescope:2021dqv}. A central challenge emerging from this progress is to understand how magnetic fields, which are ubiquitous in realistic astrophysical settings, affect black hole spacetimes and imprint themselves on observable strong field phenomena.
While in many contexts magnetic fields can be treated as environmental or perturbative effects, there also exist physically motivated solutions in which the electromagnetic field back-reacts nontrivially on the geometry and becomes an essential ingredient of the spacetime itself.

Astrophysical black holes are generically embedded in magnetized plasma environments, where magnetic fields play a crucial role in accretion dynamics~\cite{Blandford:1982xxl, Gammie:2003rj, Abramowicz:2011xu, Stuchlik:2015nlt, Hu:2025mmp, Rueda:2025lgq}, in high-energy emission processes such as synchrotron radiation and magnetic reconnection~\cite{Blandford:1977ds, Koide:2008xr, Comisso:2020ykg}, and in the launching and collimation of relativistic jets~\cite{Blandford:1982xxl, Blandford:2018iot, Saiz-Perez:2025sox}. Recent observations further indicate a close connection between jet morphology and the underlying accretion flow~\cite{Lu:2023bbn}.
In particular, long-term VLBI monitoring of M87* has revealed a measurable misalignment between the jet axis and the inferred black hole spin axis. This misalignment is accompanied by a quasi-periodic precession with a period of $11.24\pm0.47$ years and an angular velocity of $0.56\pm0.02$ rad/year~\cite{Cui:2023uyb}. Subsequent analyses indicate that this precession leaves observable imprints on both the disk size and the non-coplanar jet geometry~\cite{Cui:2024ggx}, highlighting jet precession as a promising diagnostic of black hole spacetime properties in magnetized environments.

It is generally believed that the jet and disk precession are naturally associated with the Lense-Thirring (LT) effect induced by frame dragging in rotating spacetimes~\cite{Lense:1918zz}. LT precession underlies a wide range of relativistic phenomena, including nodal orbital precession~\cite{Stone:2011mz, Everitt:2011hp, Thirring1918, LenseThirring1918, Wu:2023wld, AlZahrani:2023xix, Iorio:2024eey, Zhen:2025nah, Yuan:2025zav, Chen:2025iuu}, spin-orbit coupling in compact binaries, and the Bardeen--Petterson mechanism governing the warping and alignment of tilted accretion disks~\cite{Kopacek, Iorio:2024eey, Bardeen:1975zz, Petterson:1977JA, Papaloizou, Ostriker, Casassus, Armitage:2022six, Krolik:2015jya, Wijers, Fragile:2024osq, Drewes:2021wfa, Zhang:2025xqu, AlZahrani:2023xix}. Motivated by these insights, a phenomenological framework has recently been developed in which disk particles are modeled as following spherical orbits, the jet is assumed to originate near the disk warp radius and remain orthogonal to the disk plane, and the precession axis is identified with the black hole spin~\cite{Wei:2024cti}. This framework has been successfully applied to a variety of rotating black hole spacetimes, including Kerr~\cite{Wei:2024cti}, Kerr-Newman~\cite{Meng:2024gcg}, Kerr-Taub-NUT geometries~\cite{Meng:2025jej}, and other rotating black holes~\cite{Wang:2025fuw, Wang:2025zis, Chen:2024aaa}, yielding constraints on spin, charge, and NUT parameters from the jet precession data.

Despite these advances, a systematic understanding of how a dynamically significant electromagnetic field that back-reacts on the spacetime geometry modifies spherical orbit dynamics and LT precession remains incomplete, and the implications for parameter constraints inferred from jet observations are still poorly explored. Most existing studies either neglect magnetic fields altogether or treat them perturbatively, often at the expense of analytic control. This motivates the need for a black hole spacetime in which the electromagnetic field is incorporated at the level of the geometry while remaining analytically tractable and admitting well defined spherical orbits.

A key requirement for addressing this gap is a black hole spacetime in which a dynamically significant electromagnetic field is incorporated at the level of the geometry, while remaining analytically tractable and admitting well-defined spherical orbits. A timely opportunity is provided by the recently derived Kerr-Bertotti-Robinson (KBR) solution~\cite{Podolsky:2025tle}, which describes a rotating black hole embedded in the Bertotti-Robinson universe and permeated by a uniform Maxwell field.
The KBR spacetime interpolates smoothly between the Kerr limit ($B\to0$) and the Bertotti-Robinson limit (mass $M\to0$), while incorporating the full back reaction of the electromagnetic field on the geometry. Unlike the Kerr-Melvin spacetime~\cite{Melvin:1963qx, Melvin:1965zza, Ernst:1976mzr, Ernst:1976bsr, DiPinto:2025yaa}, the KBR solution remains algebraically type D and is asymptotically matched to a uniform electromagnetic field at large radii, rendering it a particularly suitable framework for exploring strong field phenomena in magnetized yet analytically controlled settings.

Although the KBR spacetime admits hidden symmetries and allows for the separability of null geodesics and several field equations~\cite{Gray:2025lwy, Andersson:2025bhq}, the Hamilton-Jacobi equation for timelike geodesics is generically non-separable.
This property necessitates alternative approaches to orbital dynamics and directly motivates the Hamiltonian formulation adopted in this work. Owing to these properties, the KBR geometry has rapidly emerged as a versatile arena for studies of black hole shadows~\cite{Wang:2025vsx, Zeng:2025tji, Ali:2025beh, Liu:2025wwq}, gravitational lensing~\cite{Vachher:2025jsq}, innermost stable circular orbit (ISCO) structures~\cite{Wang:2025bjf}, spinning-particle dynamics~\cite{Zhang:2025ole}, magnetic reconnection driven energy extraction~\cite{Zeng:2025olq}, magnetic Penrose process \cite{Mirkhaydarov:2026fyn}, potential implications for gravitational wave propagation~\cite{Li:2025rtf}, and the Kerr/CFT correspondence~\cite{Siahaan:2025ngu}.
Several theoretical extensions of the metric have also been proposed in Refs.~\cite{Astorino:2025lih, Podolsky:2025zlm, Ovcharenko:2025qov, Ahmed:2025ril, Ortaggio:2025sip, Barrientos:2026kdl}.

In this work, we investigate LT precession of spherical orbits in the KBR spacetime and explore how the presence of a magnetic field modifies the relation between the black hole spin, precession frequency, and disk geometry. By making use of the observed jet precession of M87*, we address the following questions: How does the magnetic parameter modify spherical orbit precession relative to the Kerr case? What constraints can be placed on the black hole spin and the warp radius of the disk in a magnetized geometry? What are the implications for the structure of the inner accretion disk? By addressing these questions, we assess the potential of jet precession observations as probes of magnetic fields in the strong gravity regime.

The paper is organized as follows. In Sec.~\ref{sec2}, we consider the spherical orbits and their stability properties in the KBR spacetime. The general expressions for the associated LT precession frequencies are presented in Sec.~\ref{sec3}. Sec.~\ref{sec4} applies the observed M87* jet precession period to constrain the black hole parameter space and the structure of the inner accretion disk. Our conclusions are summarized in Sec.~\ref{sec5}.

\section{Spherical orbits of Kerr-Bertotti-Robinson black hole}
\label{sec2}

In this section, we investigate the geodesic structure of the KBR spacetime, focusing on the spherical orbits, bound trajectories characterized by a constant radial coordinate and oscillatory motion in the polar direction. These orbits provide the natural understandings for accretion disks and constitute the dynamical foundation of the LT precession analysis developed in the next section.

\subsection{Kerr-Bertotti-Robinson black hole}
\label{sec2.1}

In this subsection, we briefly summarize those properties of the KBR geometry that are directly relevant for the geodesic dynamics analyzed in this section. The KBR solution is an exact and fully back reacted solution of the Einstein-Maxwell equations, describing a rotating black hole with mass parameter $M$ and spin parameter $a$ immersed in a large scale electromagnetic field~\cite{Podolsky:2025tle}. Unlike the test field constructions, the electromagnetic field in the KBR spacetime contributes nontrivially to the curvature and modifies the global structure of the geometry.

The KBR solution continuously interpolates between the Kerr solution ($B=0$), magnetized Schwarzschild configuration ($a=0$), and the Bertotti-Robinson universe ($M=0$). However, its asymptotic structure differs qualitatively from that of asymptotically flat black hole spacetimes.  As a result, physical properties often taken for granted in Kerr geometry, such as the existence of bound or spherical orbits extending to arbitrarily large radii, are no longer guaranteed and must be reexamined from first principles.

In Boyer-Lindquist coordinates $(t,r,\theta,\varphi)$, the line element of the KBR spacetime takes~\cite{Podolsky:2025tle}
\begin{eqnarray}
 ds^{2}=\dfrac{1}{\Omega^2}\Big[-\dfrac{{Q}}{\rho^2}
    \big(\dd t-a\sin^2\theta\,\dd\varphi\big)^2
    +\dfrac{\rho^2}{{Q}}\,\dd r^2
    +\dfrac{\rho^2}{{P}}\,\dd\theta^2\nonumber
    +\dfrac{{P}}{\rho^2}\sin^2\theta\,
    \big(a\,\dd t-(r^2+a^2)\,\dd\varphi\big)^2\,\Big],
\label{Kerr-BR}
\end{eqnarray}
where the metric functions are given by
\begin{align}
\rho^2   & = r^2+a^2\cos^2\theta\,, \label{rho2}\\
\Delta &= \Big(1-B^2M^2\,\dfrac{I_2}{I_1^2}\Big) r^2
          -2M\,\dfrac{I_2}{I_1}\,r + a^2,\label{Delta} \\
P & = 1 + B^2 \Big(M^2\,\dfrac{I_2}{I_1^2} - a^2 \Big)\cos^2\theta\,,  \label{tilde_P}\\
Q & = (1+B^2r^2)\, \Delta\,,\label{math-Q}\\
\Omega^2 & = (1+B^2r^2) - B^2 \Delta \cos^2\theta\,,\label{Omega}
\end{align}
with
\begin{align}
    I_1 = 1-\tfrac{1}{2} B^2a^2\,,\qquad
    I_2 = 1-B^2a^2\,.
\label{I1I2}
\end{align}

For later use, we list the nonvanishing components of the inverse metric
\begin{align}
 g^{tt}&=-\frac{\Omega^2}{PQ\rho^2}\left[P(r^2+a^2)^2-Qa^2\sin^2\theta\right],\quad
 g^{t\varphi}=\frac{\Omega^2a\left(Q-P(r^2+a^2)\right)}{PQ\rho^2},\\
 g^{rr}&=\frac{\Omega^2Q}{\rho^2},\quad
 g^{\theta\theta}=\frac{\Omega^2P}{\rho^2}, \quad
 g^{\varphi\varphi}=\frac{\Omega^2\left(Q-Pa^2\sin^2\theta\right)}{PQ\rho^2\sin^2\theta}.
\end{align}

The horizons of the KBR spacetime are determined by the roots of $\Delta=0$,
which yields the inner and outer horizon radii
\begin{eqnarray}
	r_{\pm}
	=
	I_1\,\frac{M I_2 \pm \sqrt{M^2 I_2-a^2 I_1^2}}{I_1^2-B^2 I_2^2}.
	\label{horizon}
\end{eqnarray}
The extremal limit corresponds to the degeneracy of the two horizons,
$r_+=r_-$, and leads to a constraint relating the spin parameter $a$,
the magnetic field strength $B$, and the mass $M$,
\begin{eqnarray}
B=\frac{\sqrt{2} \sqrt{a^2-M^2+M \sqrt{M^2-a^2}}}{a^2}.
\label{extre}
\end{eqnarray}
The condition~\eqref{extre} defines the extremality curve in the
$(a/M,\,BM)$ plane and delimits the parameter region for which the KBR spacetime admits an event horizon.

\begin{figure}[htbp]
\centering
\includegraphics[width=0.5\linewidth]{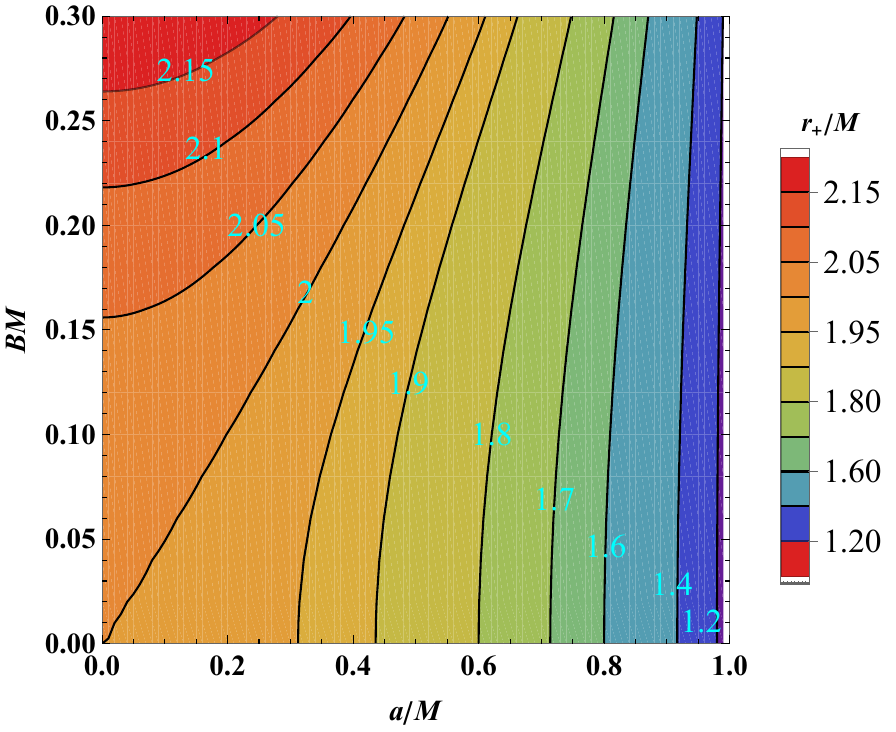}
\caption{Contour plot of the outer event horizon radius \( r_+/M \) in the KBR spacetime as a function of the dimensionless black hole spin \( a/M \) and magnetic field parameter \( BM \). The color bar indicates the value of \( r_+/M \), and the contour lines represent constant horizon radius. In the Kerr limit (\( B=0 \)), the standard Kerr horizon structure is recovered.}
\label{fig:horizoncontour}
\end{figure}

In the following analysis, we initially restrict attention to the parameter range $BM\in[0,0.3]$, which encompasses astrophysically plausible magnetic field strengths in the vicinity of supermassive black holes, and then constrain it through the observations. In Fig.~\ref{fig:horizoncontour}, we show the dependence of the outer horizon radius $r_+/M$ on the black hole spin $a/M$ and magnetic field parameter $BM$. Comparing with the Kerr case, increasing $B$ generally enlarges the horizon radius at fixed spin, reflecting the non-negligible back reaction of the electromagnetic field.

This modified horizon structure sets the innermost boundary accessible to external observers and provides an essential geometric constraint on the bound geodesic motion. As we will show in the following subsections, it plays a key role in determining the existence and stability of the spherical orbits, thereby shaping the structure of accretion disks and their associated precession dynamics.

\subsection{Geodesic motion and Hamiltonian formalism}
\label{sec2.2}

As discussed in Sec.~\ref{sec2.1}, the presence of a large-scale electromagnetic field in the KBR spacetime modifies the metric functions in a way that renders the Hamilton-Jacobi equation for timelike geodesics generically non-separable~\cite{Podolsky:2025tle, Gray:2025lwy, Wang:2025vsx}. Consequently, the standard separation of variables approach from the Kerr geometry is no longer applicable. To analyze particle motion in this non-integrable setting, we adopt the
Hamiltonian formalism, which provides a systematic and covariant
framework for geodesic dynamics and does not rely on separability.

We start with the Lagrangian of a freely falling test particle
\begin{equation}
\mathcal{L}=\frac{1}{2}g_{\mu\nu}\dot{x}^{\mu}\dot{x}^{\nu},
\end{equation}
where the overdots denote the differentiation with respect to the proper time $\tau$, and $x^{\mu}=(t,r,\theta,\varphi)$ are Boyer-Lindquist coordinates. The canonical momenta conjugate to the coordinates are defined as
\begin{equation}
p_{\mu}\equiv \frac{\partial \mathcal{L}}{\partial \dot{x}^{\mu}}
= g_{\mu\nu}\dot{x}^{\nu}.
\label{momentum}
\end{equation}
Performing the Legendre transformation yields the Hamiltonian
\begin{equation}
\mathcal{H}=p_{\mu}\dot{x}^{\mu}-\mathcal{L},
\label{Hamiltonian}
\end{equation}
which, upon using $\dot{x}^{\mu}=g^{\mu\nu}p_{\nu}$, reduces to the standard quadratic form
\begin{equation}
\mathcal{H}=\frac{1}{2}g^{\mu\nu}p_{\mu}p_{\nu}.
\label{Hamiltonian1}
\end{equation}
The equations of motion follow from Hamiltonian canonical equations
\begin{equation}
\dot{x}^{\mu}=\frac{\partial \mathcal{H}}{\partial p_{\mu}},
\qquad
\dot{p}_{\mu}=-\frac{\partial \mathcal{H}}{\partial x^{\mu}},
\label{canonicalEQ}
\end{equation}
which can be written explicitly as
\begin{align}
&\dot{t}=g^{tt}p_t+g^{t\varphi}p_\varphi,
&&\dot{p}_t=-\frac12\partial_t g^{\alpha\beta}p_\alpha p_\beta,\\
&\dot{\varphi}=g^{t\varphi}p_t+g^{\varphi\varphi}p_\varphi,
&&\dot{p}_\varphi=-\frac12\partial_\varphi g^{\alpha\beta}p_\alpha p_\beta,\\
&\dot{r}=g^{rr}p_r,
&&\dot{p}_r=-\frac12\partial_r g^{\alpha\beta}p_\alpha p_\beta,\\
&\dot{\theta}=g^{\theta\theta}p_\theta,
&&\dot{p}_\theta=-\frac12\partial_\theta g^{\alpha\beta}p_\alpha p_\beta.
\end{align}
Owing to stationarity and axisymmetry, the KBR metric is independent of $t$ and $\varphi$, implying the conservation of the corresponding conjugate momenta. These constants of motion are naturally identified with the particle’s energy $E$ and axial angular momentum $L$,
\begin{equation}
p_t=-E, \qquad p_\varphi=L.
\end{equation}
Geodesic motion is further constrained by the normalization condition
\begin{equation}
g^{\mu\nu}p_\mu p_\nu=\kappa,
\end{equation}
where $\kappa=-1$ corresponds to the timelike trajectories, which are the focus of this work. Substituting the conserved quantities into this condition and separating the radial and polar contributions yields
\begin{equation}
g^{rr}p_r^2+g^{\theta\theta}p_\theta^2
=-1-(g^{tt}E^2-2g^{t\varphi}EL+g^{\varphi\varphi}L^2).
\end{equation}
We interpret the right hand side of it as an effective potential governing the coupled radial and polar motion
\begin{equation}
\mathcal{V}_{\rm eff}(r,\theta)
=-1-(g^{tt}E^2-2g^{t\varphi}EL+g^{\varphi\varphi}L^2),
\end{equation}
so that the allowed region of motion satisfies
\begin{equation}
g^{rr}p_r^2+g^{\theta\theta}p_\theta^2
=\mathcal{V}_{\rm eff}(r,\theta)\ge 0.
\label{rpmotion}
\end{equation}
Outside the event horizon, this condition restricts the allowed phase space and identifies turning points in the radial or polar directions through the zeros of $\mathcal{V}_{\rm eff}$.

This Hamiltonian formulation provides the natural starting point for identifying the spherical orbits, which correspond to the special solutions satisfying both radial equilibrium and bounded polar motion. These orbits are analyzed in detail below.

\subsection{Spherical orbits}
\label{sec2.3}

Spherical orbits form a distinguished class of bound geodesics in stationary and axisymmetric spacetimes. They are characterized by a constant radial coordinate while allowing oscillatory motion in the polar direction. In contrast to the equatorial circular orbits, which are confined to the plane $\theta=\pi/2$, spherical orbits occupy a two dimensional surface specified by a fixed radius and a pair of symmetric polar turning points. Owing to this geometry, spherical orbits provide a natural dynamical framework for modeling tilted accretion disks, where the orbital angular momentum of the disk is misaligned with the black hole spin axis. In this subsection, we extend the theory of spherical orbits to the KBR spacetime within the Hamiltonian framework introduced above.

From the Hamiltonian formulation developed in Sec.~\ref{sec2.2}, the coupled radial and polar motions are governed by Eq.~\eqref{rpmotion}, where the effective potential $\mathcal{V}_{\rm eff}(r,\theta)$ depends on the conserved energy $E$ and axial angular momentum $L$. Within this framework, a spherical orbit is realized as a special class of bound solutions characterized by a constant radial coordinate $r=r_0$ and bounded polar oscillations between two symmetric turning points
\begin{equation}
\theta_{\rm t}=\frac{\pi}{2}\pm\zeta,
\end{equation}
where $\zeta$ denotes the inclination angle with respect to the equatorial plane. At each polar turning point, the polar momentum vanishes
\begin{equation}
p_{\theta}\big|_{\theta=\theta_{\rm t}}=0,
\label{turn_theta}
\end{equation}
while radial equilibrium requires
\begin{equation}
p_{r}=0, \qquad \dot{p}_{r}=0,
\label{turn_r}
\end{equation}
both evaluated at $r=r_0$. The condition $\dot{p}_r=0$ ensures the stationarity of the radial motion and serves as the defining dynamical criterion for a spherical orbit.

Imposing these requirements on the Hamiltonian constraint leads to
\begin{equation}
\mathcal{V}_{\rm eff}(r_0,\theta_{\rm t})=0,
\label{turn}
\end{equation}
which ensures that the orbit lies on the boundary of the allowed region in phase space. Together with the radial equilibrium condition \eqref{turn_r}, this yields two algebraic relations evaluated at $\theta_{\rm t}$ and $r_0$,
\begin{align}
& \left(E^2\partial_r g^{tt}-2EL \partial_r g^{t\varphi}
+L^2 \partial_r g^{\varphi\varphi}\right)=0,
\label{cond1}\\
& -1-(E^2g^{tt}-2ELg^{t\varphi}
+L^2g^{\varphi\varphi})=0.
\label{cond2}
\end{align}
Solving Eqs.~\eqref{cond1}-\eqref{cond2}, one can determine the conserved energy $E_s$ and angular momentum $L_s$ associated with spherical orbits~\cite{Meng:2024gcg},
\begin{align}
&E_s=\frac{1}{\sqrt{-g^{tt}-2g^{t\varphi}\chi
+g^{\varphi\varphi}\chi^{2}}}
\Big|_{\theta=\theta_{\rm t},r=r_0},
\label{Es}\\
&L_s=\frac{-\chi}{\sqrt{-g^{tt}-2g^{t\varphi}\chi
+g^{\varphi\varphi}\chi^{2}}}
\Big|_{\theta=\theta_{\rm t},r=r_0},
\label{Ls}
\end{align}
where the ratio $\chi=p_{\varphi}/p_{t}$ is given by
\begin{equation}
\chi=\frac{-\partial_r g^{t\varphi}\pm
\sqrt{(\partial_r g^{t\varphi})^2
-\partial_r g^{tt}\partial_r g^{\varphi\varphi}}}
{\partial_r g^{\varphi\varphi}}
\Big|_{\theta=\theta_{\rm t},r=r_0},
\end{equation}
with the upper (lower) sign corresponding to the prograde (retrograde) spherical orbits. In the appropriate limits, these expressions reduce smoothly to the well known Kerr results, while fully encoding the electromagnetic back reaction characteristic of the KBR geometry.

Once $E_s$ and $L_s$ are fixed for a given inclination angle $\zeta$, the complete spacetime trajectory of a spherical orbit follows from integrating the remaining first-order Hamilton equations
\begin{align}
&\dot{t}= -g^{tt}E_s + g^{t\varphi} L_s, \\
&\dot{\varphi}= -g^{t\varphi}E_s + g^{\varphi\varphi} L_s, \\
&\dot{\theta}= g^{\theta\theta} p_{\theta}, \\
&\dot{p}_{\theta}= -\tfrac12\left(
E_s^{2}\partial_{\theta}g^{tt}
-2E_sL_s\partial_{\theta}g^{t\varphi}
+L_s^{2}\partial_{\theta}g^{\varphi\varphi}
+p_{\theta}^{2}\partial_{\theta}g^{\theta\theta}
\right).
\label{soevo}
\end{align}

\begin{figure}[htbp]
\center{\subfigure[ ]{\label{Erpro}
\includegraphics[width=5cm]{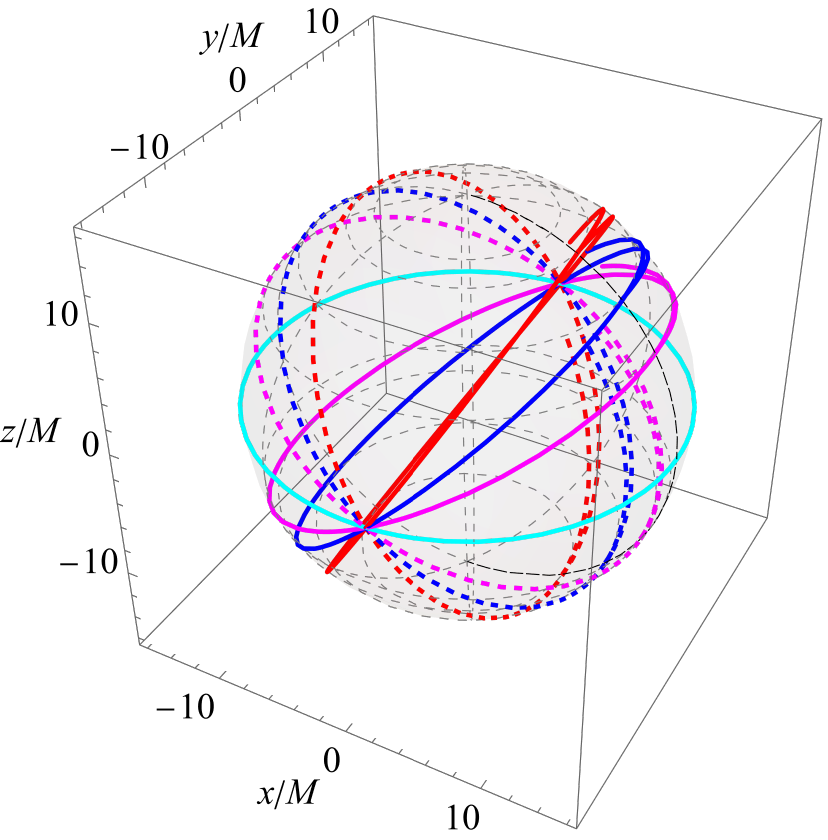}}
\subfigure[ ]{\label{Lrpro}
\includegraphics[width=5cm]{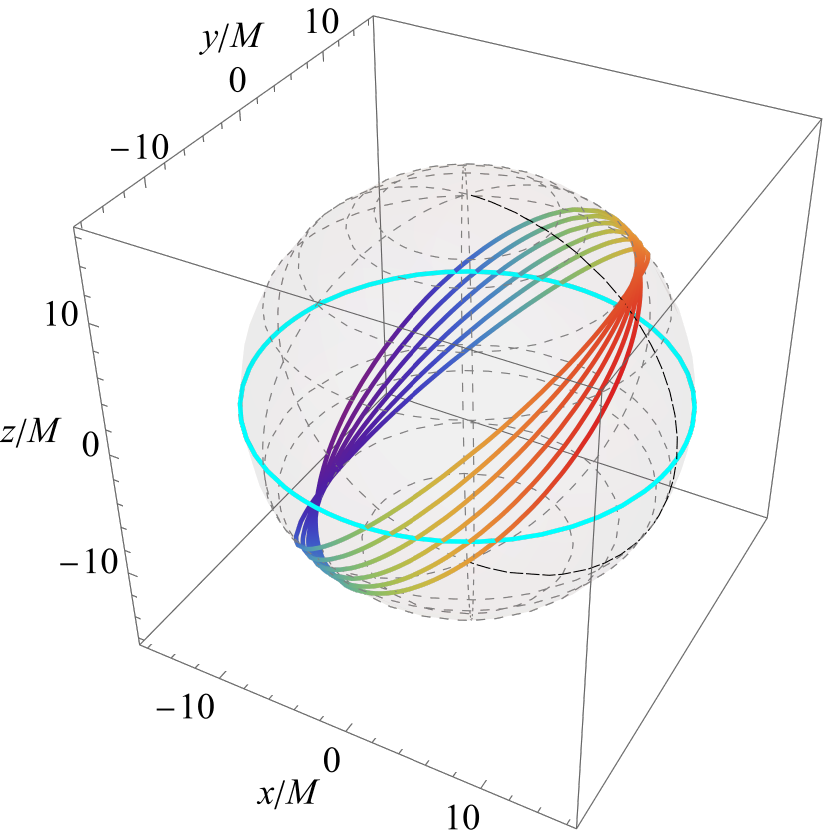}}\\
\subfigure[ ]{\label{Erret}
\includegraphics[width=5cm]{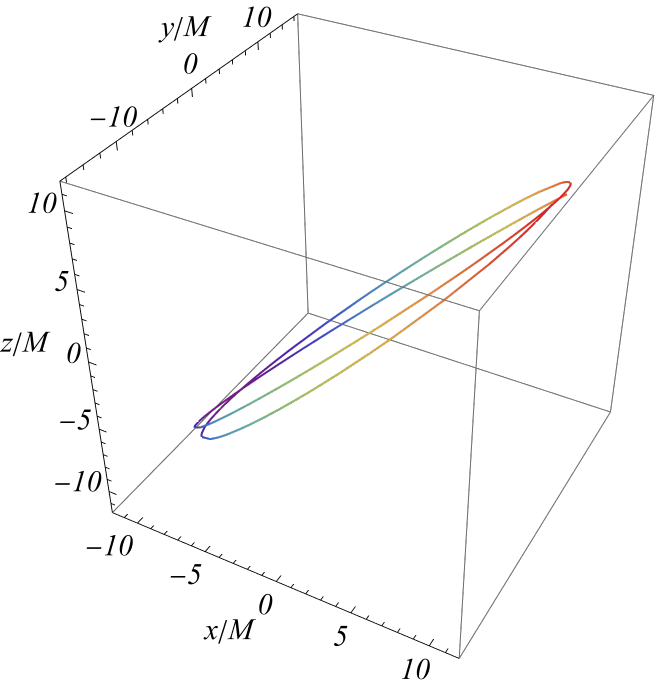}}
\subfigure[ ]{\label{Lrret}
\includegraphics[width=5cm]{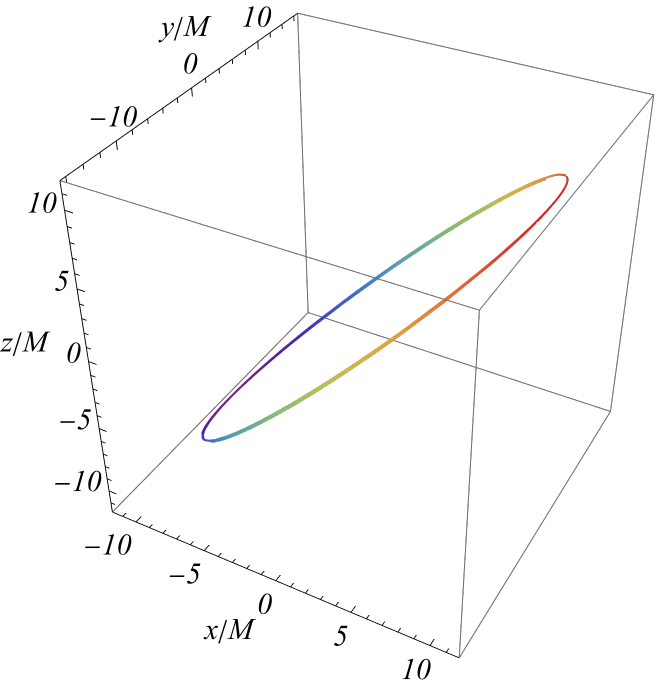}}}
\caption{
Visualization of spherical orbits and their LT precession in the KBR spacetime for fixed parameters $a/M=0.9$ and $BM=0.01$. (a) Families of the spherical orbits with different inclination angles $\zeta=\pi/6,\ \pi/4,\ \pi/3$ (in magenta, blue, and red colors), shown together with the equatorial cyan ring. Solid (dashed) curves correspond to the prograde (retrograde) motions. (b) A representative inclined spherical orbit with $\zeta=\pi/4$, illustrating the secular precession of the orbital plane about the spin axis when described with respect to the. Boyer-Lindquist coordinates. (c) Projection of the same orbit onto the plane perpendicular to the spin axis, showing a non-closed trajectory due to the cumulative LT precession. (d) The same trajectory represented in a uniformly rotating reference frame with angular velocity equal to the mean precession rate, in which the underlying closed trajectory is recovered. These figures illustrate the geometric origin of the precession frequency $\omega_{\rm LT}$ defined in Sec.~\ref{sec3}.}
\label{tra}
\end{figure}

To illustrate the three dimensional geometry and precession properties for the spherical orbits, the representative trajectories are shown in Fig.~\ref{tra} for a KBR background with fixed parameters $a/M=0.9$, $BM=0.01$. In Fig. \ref{Erpro}, we display the spherical orbits with inclination angles $\zeta=\pi/6,\ \pi/4,\ \pi/3$ (in magenta, blue, and red colors), together with the equatorial cyan ring. The solid and dashed curves are for the prograde and retrograde motions, respectively, demonstrating how the orbital plane tilts as the inclination increases. Fig. \ref{Lrpro} shows the evolution of a single inclined spherical orbit with $\zeta=\pi/4$, revealing the secular precession of the orbital plane about the black hole spin axis. Obviously, in the Boyer-Lindquist coordinates, the trajectory does not close after one polar oscillation. Figs. \ref{Erret} and \ref{Lrret} further clarify the geometric nature of this precession. In Fig. \ref{Erret}, it shows the projection of the orbit onto the plane perpendicular to the spin axis, where the cumulative azimuthal advance leads to a non-closed trajectory. In Fig. \ref{Lrret}, the same orbit is represented in a uniformly rotating reference frame that subtracts the mean precession rate, revealing the underlying closed trajectory. This comparison clearly highlights the interpretation of LT precession as a uniform rotation of an otherwise closed spherical orbit.

\begin{figure}[htbp]
\centering
\subfigure[Prograde spherical orbits]{\label{ELpro}
\includegraphics[width=0.43\linewidth]{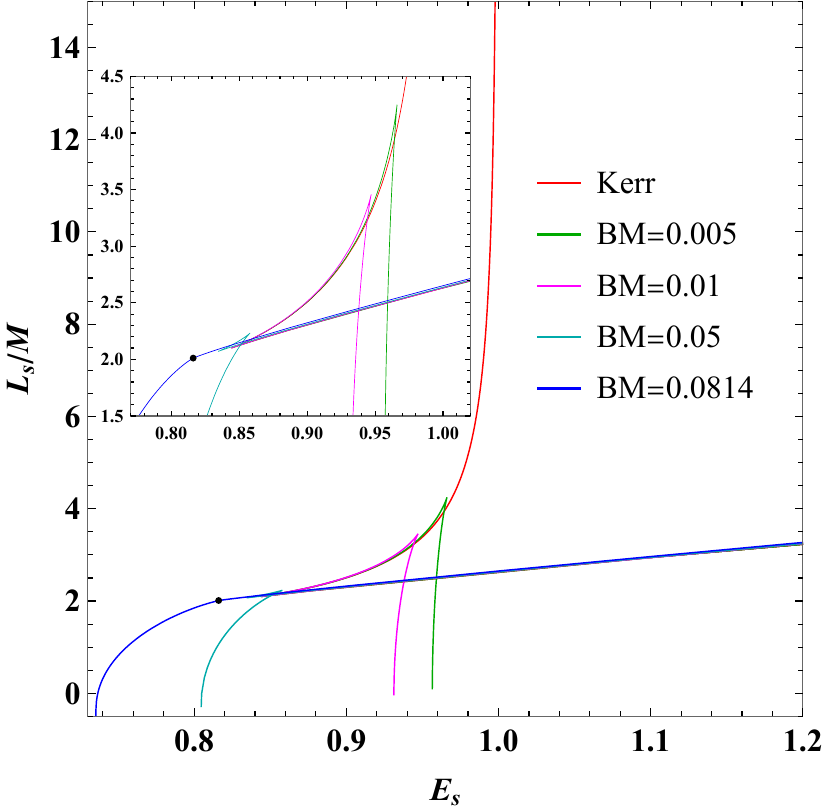}}
\subfigure[Retrograde spherical orbits]{\label{ELret}
\includegraphics[width=0.45\linewidth]{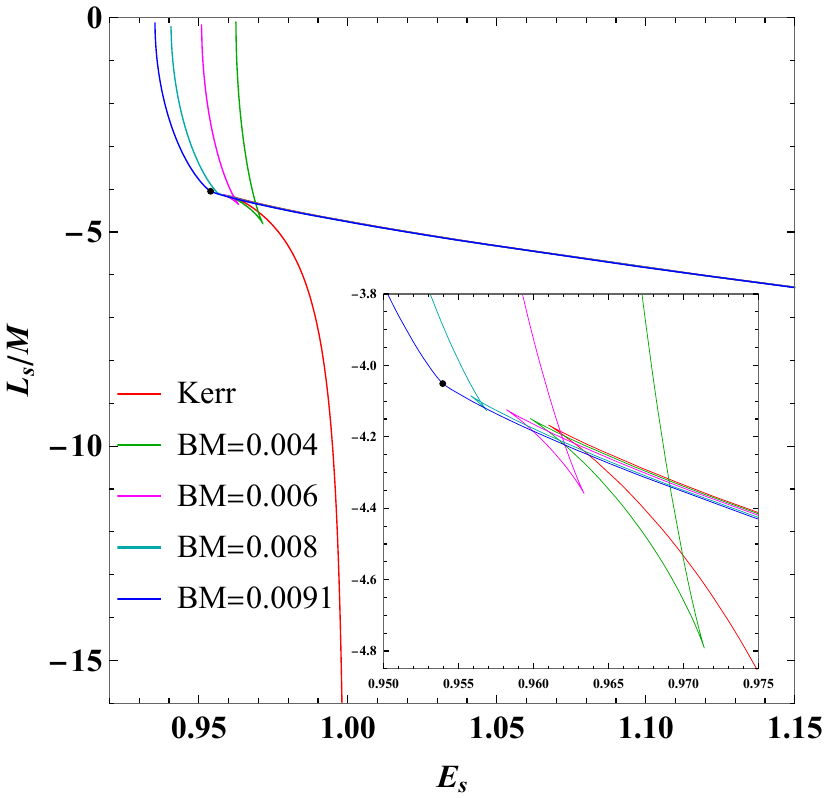}}
\caption{Energy-angular momentum ($E_s$, $L_s$) diagrams for the spherical orbits
in the KBR spacetime with fixed spin $a/M=0.9$ and inclination angle
$\zeta=1.25^\circ$. Each curve is. parametrized by the orbital radius $r_0$.
The curves originate near the outer event horizon at $E_s>1$ and evolve
as $r_0$ increases. (a) Prograde spherical orbits. (b) Retrograde spherical orbits.
}
\label{spoEL}
\end{figure}

Having established the kinematic structure of the spherical orbits, we now examine their representation in the energy-angular momentum parameter space.
In Fig.~\ref{spoEL}, we display the parametric curves $E_s(r_0)$-$L_s(r_0)$ for the spherical orbits at fixed black hole spin $a/M=0.9$ and inclination angle $\zeta=1.25^\circ$ with the orbital radius $r_0$ serving as the parameter. In both the prograde and retrograde cases, the curves originate on the right hand side of the diagram at $E_s>1$, corresponding to the spherical orbits located just outside the outer event horizon, and evolve continuously in the $E_s$-$L_s$ space as $r_0$ increases.

In the Kerr case ($B=0$), it exhibits a single cusp. As $r_0$ increases beyond this point, the curve extends smoothly toward $E_s\to1$, with the angular momentum increasing and decreasing monotonically for prograde and retrograde orbits. For this case, the spherical orbits persist to arbitrarily large radii, and the orbital angular momentum $L_s$ can grow without bound as $r_0\to\infty$, reflecting the asymptotically flat nature of the Kerr spacetime. The cusp therefore marks the unique transition between radially unstable and stable spherical motion~\cite{Meng:2024gcg}.

When a finite magnetic field is present, the evolution of the $E_s$-$L_s$ curves changes qualitatively. With the increasing of $r_0$, the angular momentum initially decreases as the energy decreases, until the curve reaches an inner cusp. As $r_0$ increases further, the curve reverses direction and both $E_s$ and $L_s$ increase, giving rise to an intermediate branch. Upon further increasing $r_0$, a second cusp appears, beyond which, the angular momentum decreases again. In contrast to the Kerr case, the curve does not extend to arbitrarily large radius. Instead, as $r_0$ approaches a maximal value, the orbital angular momentum $L_s$ tends to zero while the energy remains below unity, $E_s<1$. The $E_s$-$L_s$ curve therefore terminates at a finite radius, highlighting a sharp qualitative difference from the Kerr black hole case. As a result, the $E_s$-$L_s$ curve acquires a characteristic
swallowtail behavior with two cusps, signaling a nontrivial reorganization of spherical orbits in the presence of the magnetic field.

A detail study shows that the extent of the swallowtail region depends sensitively on the
magnetic field strength. For smaller values of $B$, the swallowtail encloses a larger region in the ($E_s$, $L_s$) parameter space. While increasing $B$ causes the two cusps approach each other and the swallowtail behavior shrinks. At a critical magnetic field $B=B_{\mathrm{cri}}$, the two cusps merge into a single point, beyond which the energy $E_s$ becomes a monotonic function of $L_s$. Hence, this magnetic field driven behavior marks a qualitative transitionin the structure of the spherical orbit parameter space.

These topological features provide clear geometric signatures of changes in radial stability and imply the existence of several distinguished spherical orbits. Their precise identification, stability properties, and associated characteristic radii are analyzed systematically in the following subsection.

\subsection{Innermost stable spherical orbits, marginally stable spherical orbits, and critical spherical orbits}
\label{sec2.4}

The analysis of the spherical orbits in ($E_s$, $L_s$) parameter space presented in
Fig.~\ref{spoEL} reveals nontrivial topological structures, cusps and swallowtail configurations, encoding qualitative changes in the radial stability. In the KBR spacetime, these geometric features are not merely kinematic curiosities but provide a direct and compact representation of how stable and unstable spherical orbits are structured in parameter space. In this subsection, we make this correspondence explicitly by establishing a systematic mapping between the topology of the $E_s$-$L_s$ curves and the radial stability properties of the spherical orbits.

Before turning to the explicit radial dependence of $E_s(r_0)$ and $L_s(r_0)$, it is instructive to clarify how the topological structures observed in the $(E_s,L_s)$ space (Fig.~\ref{spoEL}) are encoded in the single valued functions $E_s(r_0)$ and $L_s(r_0)$. Since each point on the $E_s$-$L_s$ curve corresponds to a spherical
orbit at a specific radius $r_0$, cusp structures in Fig.~\ref{spoEL} map
directly to extrema of $E_s(r_0)$ or $L_s(r_0)$, where $dE_s/dr_0=0$ or $dL_s/dr_0=0$. Conversely, the number of spherical orbits admitted for a given $(E_s,L_s)$ pair is determined by the number of radial solutions $r$ associated with that point on the curve, thereby encoding the radial stability properties in a purely geometric manner.

For a fixed inclination angle $\zeta$, a spherical orbit is uniquely labeled by its orbital radius $r_0$. Its stability against small radial perturbations is determined by the
local structure of the effective radial potential governing the radial motion. Within the Hamiltonian framework adopted here, this stability criterion can be expressed equivalently in terms of the radial dependence of the conserved energy and angular momentum of the spherical orbits, $E_s(r_0)$ and $L_s(r_0)$ \cite{Meng:2024gcg, Meng:2025jej}. A detailed derivation and supporting analysis of this correspondence can be found in Ref.~\cite{Meng:2025jej}. In particular, a radially stable spherical orbit corresponds to a local minimum of $E_s(r_0)$ at fixed $L_s$, or, equivalently, to a local minimum of $L_s(r_0)$ at fixed $E_s$. Marginally stable spherical orbits occur at extrema where the first radial derivative vanishes and the second derivative changes sign. These points therefore mark the boundaries between stable and
unstable radial motion.

\begin{figure}[htbp]
\centering
\subfigure[$E_s$ vs. $r_0/M$ for prograde orbits]{\label{Erproa}
\includegraphics[width=0.45\linewidth]{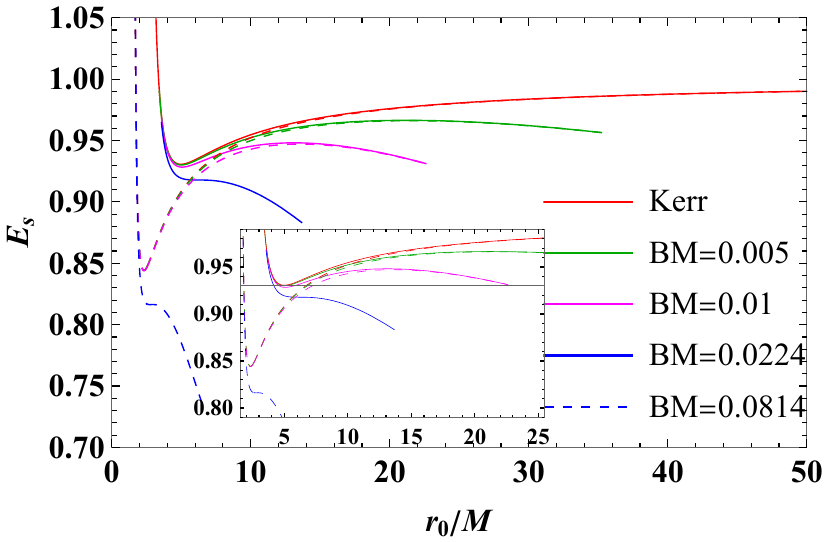}}
\subfigure[$L_s/M$ vs. $r_0/M$ for prograde orbits]{\label{Lrproa}
\includegraphics[width=0.45\linewidth]{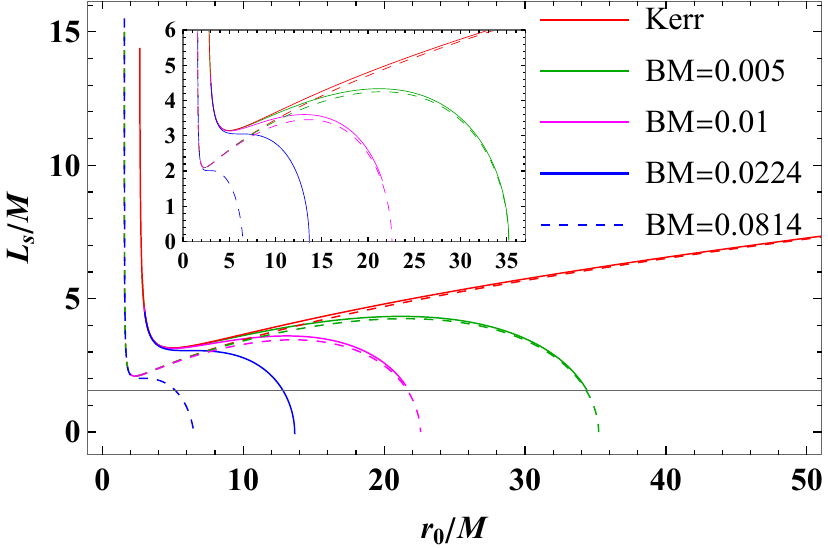}}
\\
\subfigure[$E_s$ vs. $r_0/M$ for retrograde orbits]{\label{Erreta}
\includegraphics[width=0.45\linewidth]{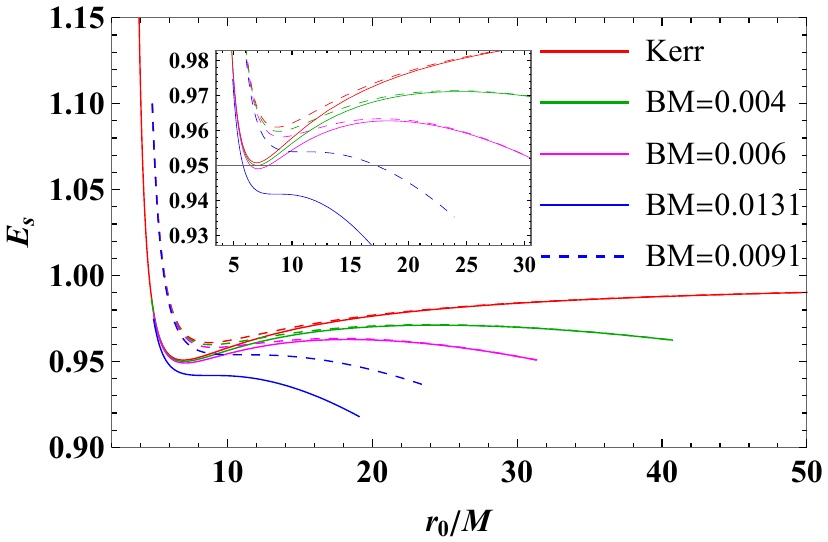}}
\subfigure[$L_s/M$ vs. $r_0/M$ for retrograde orbits]{\label{Lrreta}
\includegraphics[width=0.45\linewidth]{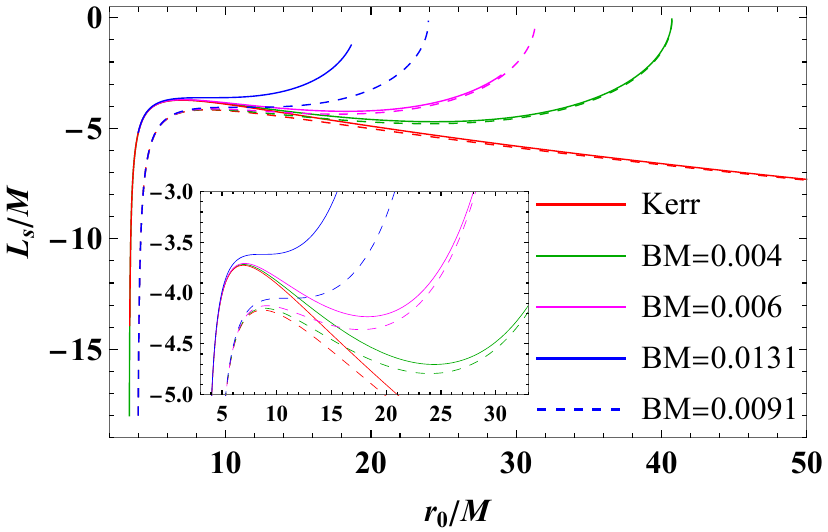}}
\caption{
Energy and angular momentum of spherical orbits as functions of the orbital radius $r_0/M$ in the KBR spacetime with the inclination angle $\zeta = 1.25^\circ$.
The upper (lower) row shows prograde (retrograde) orbits.
Solid and dashed curves represent black hole spins $a/M = 0.3$ and $0.9$, respectively, while different colors denote different magnetic field strengths $B$ as indicated. Local extrema of $E_s(r_0)$ or $L_s(r_0)$ signal transitions in radial stability and identify characteristic spherical orbits, including the ISSOs and the MSSOs.
These extrema provide the dynamical origin of the cusp and swallowtail structures observed in the $(E_s, L_s)$ space shown in Fig.~\ref{spoEL}. (a) $E_s(r_0)$ for prograde orbits. (b) $L_s(r_0)$ for prograde orbits. (c) $E_s(r_0)$ for retrograde orbits. (d) $L_s(r_0)$ for retrograde orbits.
}
\label{srEl}
\end{figure}

In Fig.~\ref{srEl}, taking two representative spin $a/M=0.3$ and $0.9$, we show $E_s(r_0)$ and $L_s(r_0)$ as a function of $r_0$ for prograde (upper row) and retrograde (lower row) spherical orbits at fixed inclination $\zeta=1.25^\circ$, while varying magnetic field strength $BM$. These curves provide a direct and transparent mapping between extrema in $E_s(r_0)$ or $L_s(r_0)$ and the cusp structures identified in the $(E_s,L_s)$ space shown in Fig.~\ref{spoEL}. This correspondence can be made explicitly by comparing the Kerr and KBR cases.
In the Kerr spacetime ($B=0$), the single cusp in Fig.~\ref{spoEL}
corresponds to the unique extremum of $E_s(r_0)$ or $L_s(r_0)$, identified
as the innermost stable spherical orbit. For a fixed value of $E_s$ or $L_s$ below the cusp, two spherical orbits exist at most: a smaller radius branch is radially unstable and a larger radius branch is radially stable. Accordingly, the branch of the $E_s$-$L_s$ curve beyond the cusp represents the stable spherical orbits, while the branch preceding the cusp corresponds to the unstable ones.

In contrast, for the KBR spacetime with $0<B<B_{\mathrm{cri}}$, the swallowtail structure in Fig.~\ref{spoEL} implies that, for a given value of $E_s$ or $L_s$, up to three distinct spherical orbits may coexist. Among these, only the orbit associated with the intermediate branch of the swallowtail is radially stable. In the radial representation, this stable branch corresponds precisely to the interval $r_{\mathrm{ISSO}}<r_0<r_{\mathrm{MSSO}}$, while spherical orbits with $r_0<r_{\mathrm{ISSO}}$ or $r_0>r_{\mathrm{MSSO}}$ are radially unstable.

Guided by this correspondence, we identify three distinct classes of characteristic spherical orbits that govern the stability and existence of spherical motion.

\begin{enumerate}
\item \textbf{Innermost stable spherical orbit (ISSO).}
This orbit is defined as the stable spherical orbit with the smallest radius and marks the inner boundary of radially stable motion. At $r_0=r_{\mathrm{ISSO}}$, the conserved energy or angular momentum exhibits a local minimum
\begin{eqnarray}
\left.\frac{d E_s}{dr_0}\right|_{r_0=r_{\mathrm{ISSO}}}=0,
\qquad
\left.\frac{d^2 E_s}{dr_0^2}\right|_{r_0=r_{\mathrm{ISSO}}}>0,
\label{eq:ISSO_E}
\end{eqnarray}
or, equivalently,
\begin{eqnarray}
\left.\frac{d L_s}{dr_0}\right|_{r_0=r_{\mathrm{ISSO}}}=0,
\qquad
\left.\frac{d^2 L_s}{dr_0^2}\right|_{r_0=r_{\mathrm{ISSO}}}>0.
\label{eq:ISSO_L}
\end{eqnarray}
Inside $r_{\mathrm{ISSO}}$, all spherical orbits are radially unstable. In Fig.~\ref{srEl}, the ISSO corresponds to the innermost extremum of $E_s(r_0)$ or $L_s(r_0)$ with positive curvature.

\item \textbf{Marginally stable spherical orbit (MSSO).}
In the KBR spacetime, the presence of the external magnetic field can generate additional extrema in $E_s(r_0)$ and $L_s(r_0)$ beyond the ISSO. The outer extremum, which is a local maximum, defines a second marginal stability point,
\begin{eqnarray}
\left.\frac{d E_s}{dr_0}\right|_{r_0=r_{\mathrm{MSSO}}}=0,
\qquad
\left.\frac{d^2 E_s}{dr_0^2}\right|_{r_0=r_{\mathrm{MSSO}}}<0,
\label{eq:MSSO_E}
\end{eqnarray}
or, equivalently,
\begin{eqnarray}
\left.\frac{d L_s}{dr_0}\right|_{r_0=r_{\mathrm{MSSO}}}=0,
\qquad
\left.\frac{d^2 L_s}{dr_0^2}\right|_{r_0=r_{\mathrm{MSSO}}}<0.
\label{eq:MSSO_L}
\end{eqnarray}
Radial stability is therefore restricted to the finite interval $r_{\mathrm{ISSO}}<r_0<r_{\mathrm{MSSO}}$. The MSSO corresponds to the second cusp in the swallowtail structure of Fig.~\ref{spoEL} and represents a genuinely magnetic field induced feature that is absent in the asymptotically flat Kerr spacetime.

\item \textbf{Critical spherical orbit (CSO) and critical magnetic field.}
As the magnetic field strength increases, the ISSO and MSSO move toward each other. At a critical value $B=B_{\mathrm{cri}}(a,\zeta)$, the two extrema merge into a single inflection point,
\begin{eqnarray}
\left.\frac{d E_s}{dr_0}\right|_{r_0=r_{\mathrm{CSO}}}=0,
\qquad
\left.\frac{d^2 E_s}{dr_0^2}\right|_{r_0=r_{\mathrm{CSO}}}=0,
\label{eq:criSO_E}
\end{eqnarray}
with an equivalent formulation in terms of $L_s(r_0)$. The corresponding orbit defines the critical spherical orbit. For $B>B_{\mathrm{cri}}$, the functions $E_s(r_0)$ and $L_s(r_0)$ become monotonic. The swallowtail structure disappears, and both the ISSO and MSSO cease to exist.
\end{enumerate}

\begin{figure}[htbp]
\centering
\subfigure[$r_{\mathrm{ISSO}}/M$ of prograde orbits]{\label{risso_pro}
\includegraphics[width=0.45\linewidth]{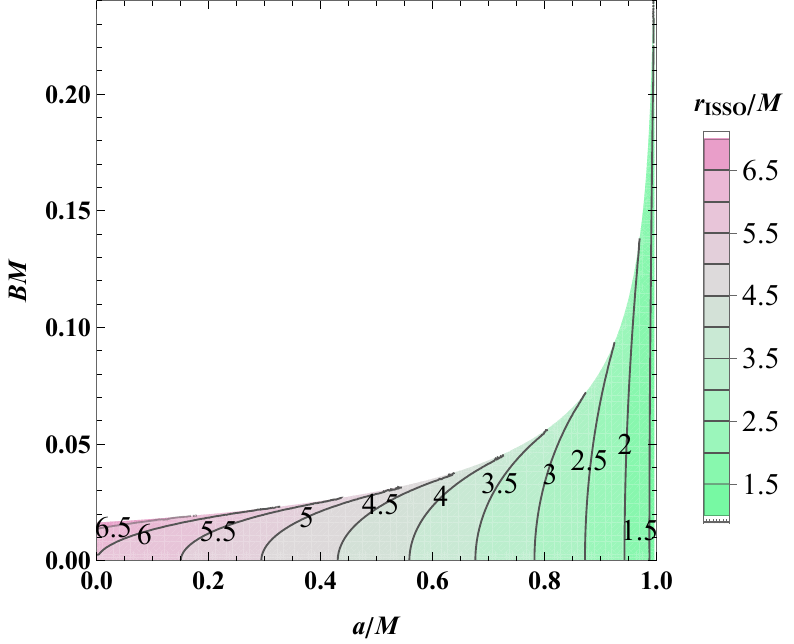}}
\subfigure[$r_{\mathrm{ISSO}}/M$ of retrograde orbits]{\label{risso_ret}
\includegraphics[width=0.45\linewidth]{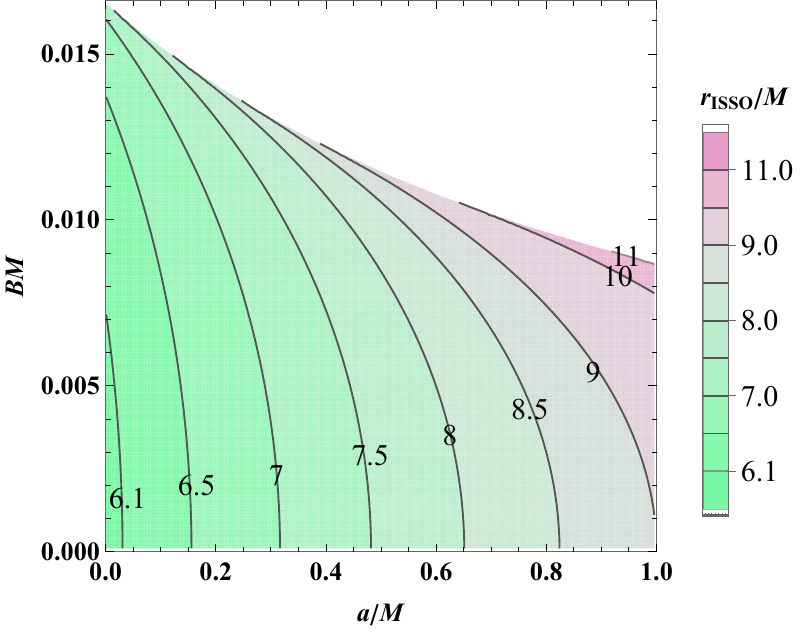}}
\caption{
Contour plots of the radius $r_{\mathrm{ISSO}}/M$ of ISSOs in the $(a/M, BM)$ parameter space for a fixed inclination angle $\zeta = 1.25^\circ$. The color bar indicates the dimensionless radius $r_{\mathrm{ISSO}}/M$. These contours illustrate how the location of the ISSOs varies across the parameter space and provide the lower bound for the warp radius $r_{\mathrm{w}}$ used in the disk precession analysis of Sec.~\ref{sec4}. (a) Prograde orbits. (b) Retrograde orbits.}
\label{risso}
\end{figure}

\begin{figure}[htbp]
\centering
\subfigure[$r_{\mathrm{MSSO1}}/M$ of prograde orbits with $BM \in (0.003, 0.24)$]{\label{rmsso1_pro}
\includegraphics[width=0.45\linewidth]{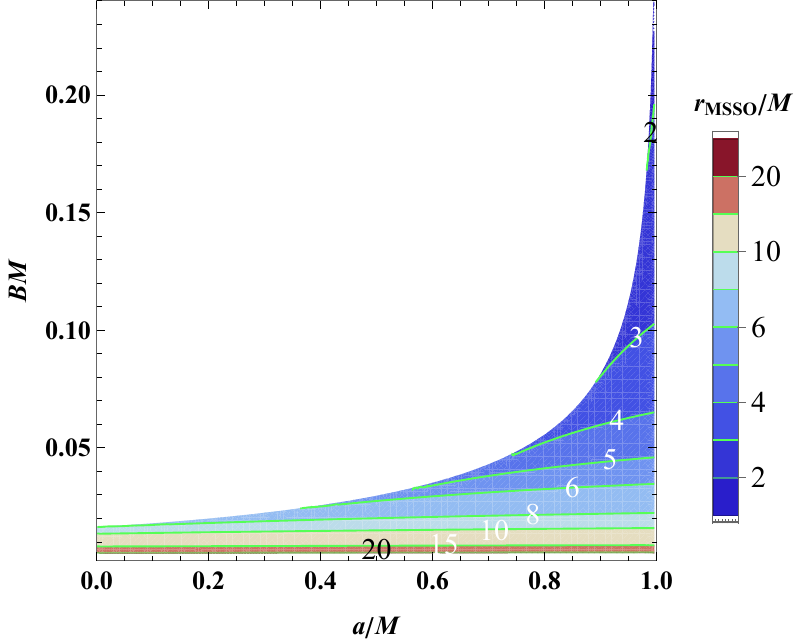}}
\subfigure[$r_{\mathrm{MSSO1}}/M$ of retrograde orbit with $BM \in (0.003, 0.0165)$]{\label{rmsso1_ret}
\includegraphics[width=0.45\linewidth]{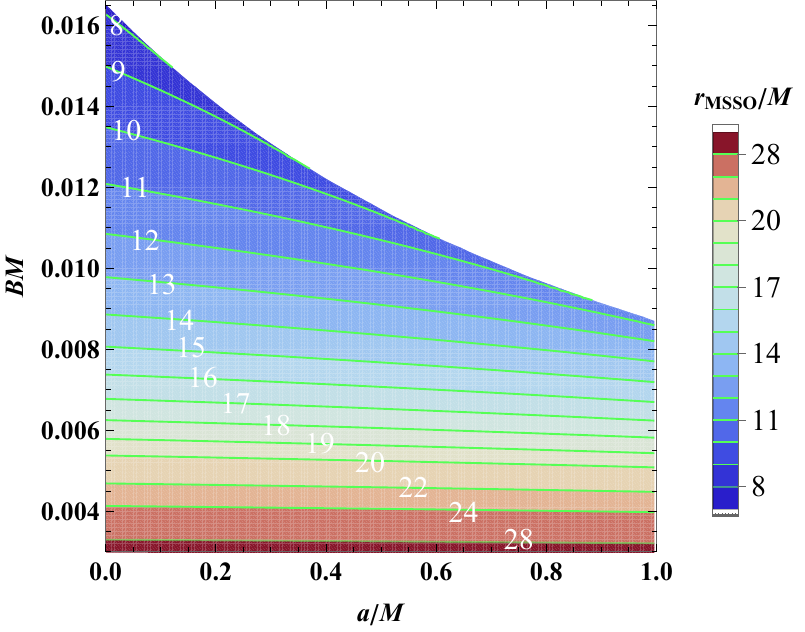}}
\\
\subfigure[$r_{\mathrm{MSSO2}}/M$ of prograde orbits with $BM \in (0.00001, 0.015)$]{\label{rmsso2_pro}
\includegraphics[width=0.45\linewidth]{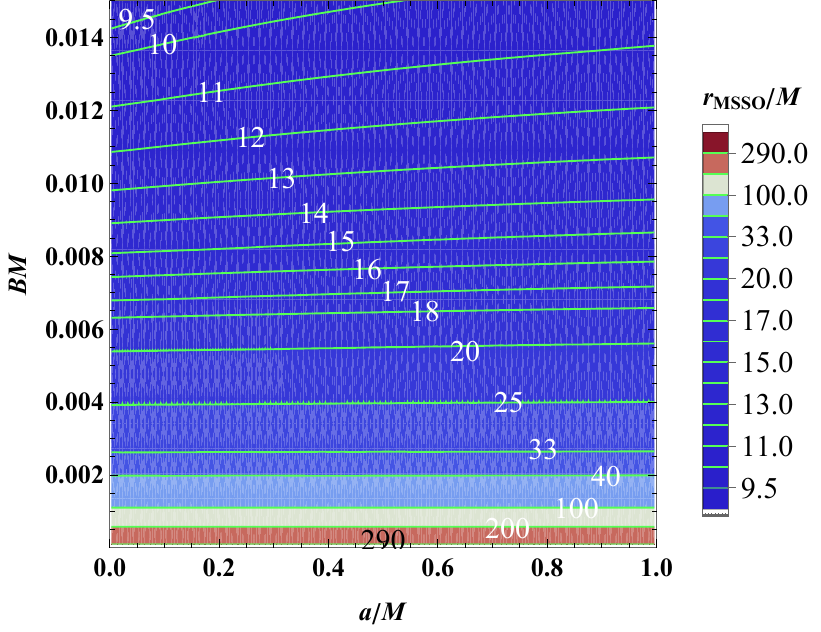}}
\subfigure[$r_{\mathrm{MSSO2}}/M$ of retrograde orbits with $BM \in (0.00001, 0.015)$]{\label{rmsso2_ret}
\includegraphics[width=0.45\linewidth]{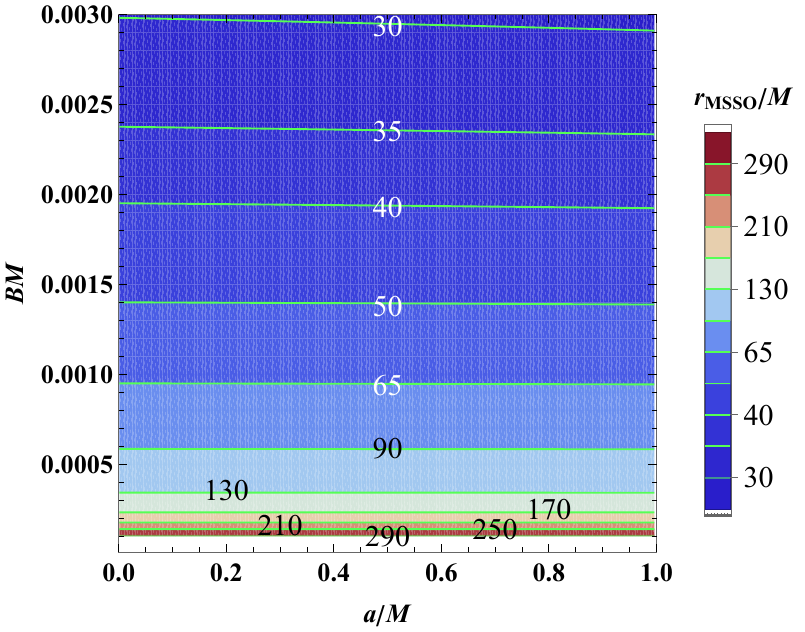}}
\caption{
Contour plots of the radius $r_{\mathrm{MSSO}}/M$ of the MSSOs in the $(a/M, BM)$ parameter space for a fixed inclination $\zeta = 1.25^\circ$.
To accommodate the wide dynamical range of $r_{\mathrm{MSSO}}$, the MSSOs are displayed in two branches:
$r_{\mathrm{MSSO1}}$ (first row), corresponding to moderate and strong magnetic fields, and $r_{\mathrm{MSSO2}}$ (second row), covering the weak field regime.
These branches exist only below the critical line $B_{\mathrm{cri}}(a)$ (see Fig.~\ref{criso}), where they merge to form the critical spherical orbit $r_{\mathrm{CSO}}$.
(a) $r_{\mathrm{MSSO1}}/M$ of prograde orbits with $BM \in (0.003, 0.24)$. (b) $r_{\mathrm{MSSO1}}/M$ of retrograde orbit with $BM \in (0.002, 0.0165)$. (c) $r_{\mathrm{MSSO2}}/M$ of prograde orbits with $BM \in (0.0001, 0.003)$. (d) $r_{\mathrm{MSSO2}}/M$ of retrograde orbits with $BM \in (0.0001, 0.002)$.
}
\label{rimsso}
\end{figure}

\begin{figure}[htbp]
\centering
\subfigure[$BM$ vs. $a/M$ for prograde critical spherical orbits]{\label{ELprohs}
\includegraphics[width=0.44\linewidth]{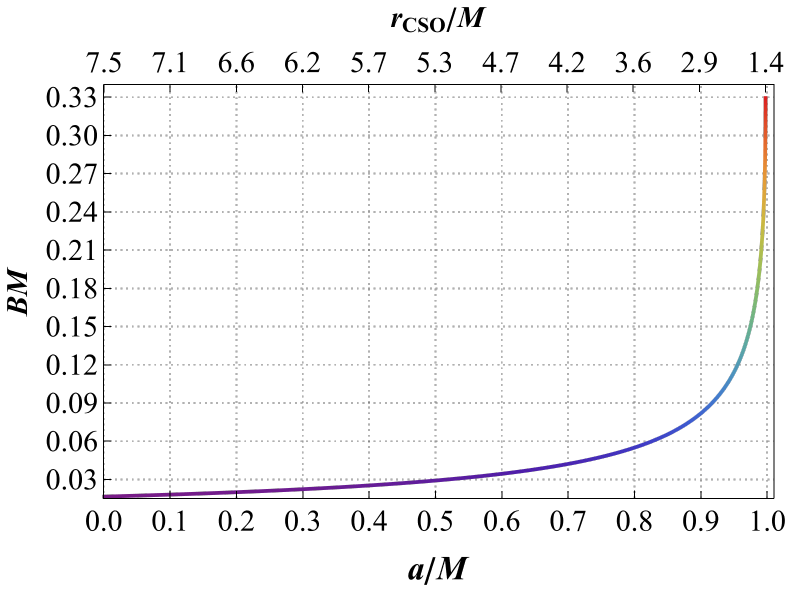}}
\subfigure[$BM$ vs. $a/M$ for retrograde critical spherical orbits]{\label{ELreths}
\includegraphics[width=0.45\linewidth]{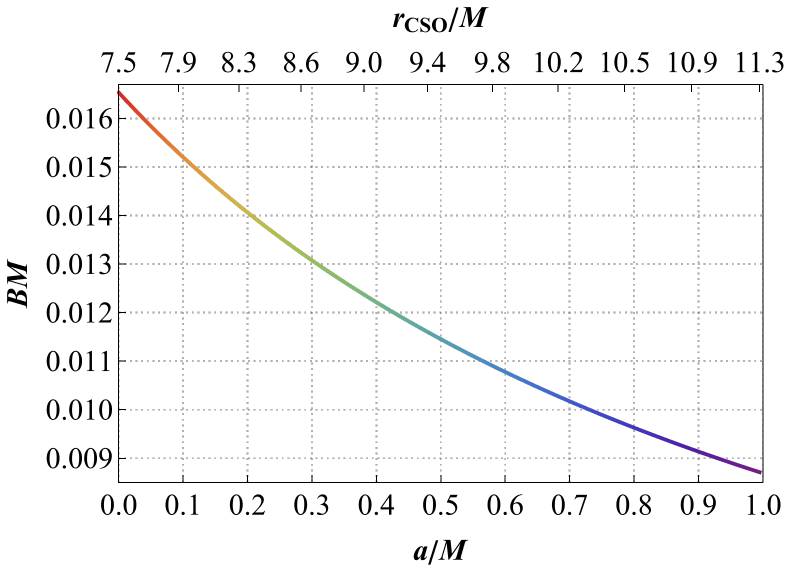}}
\caption{Radius $r_{\mathrm{CSO}}/M$ of the critical spherical orbits encoded by color as a function of the black hole spin $a/M$ and the critical magnetic field $B_{\mathrm{cri}M}$ at $\zeta = 1.25^\circ$.
Each curve represents the locus of critical points $\{a_{\mathrm{cri}}/M, B_{\mathrm{cri}}M, r_{\mathrm{CSO}}/M\}$ at which the swallowtail structure in the $(E_s,L_s)$ parameter space disappears. Regions with $B < B_{\mathrm{cri}}$ admit swallowtail structures, whereas for $B > B_{\mathrm{cri}}$ the spherical orbit energy-angular momentum relation becomes single valued. (a) Prograde spherical orbits. (b) retrograde spherical orbits.
}
\label{criso}
\end{figure}

The dependence of these characteristic radii on the black hole parameters
$(a/M,BM)$ is summarized in Figs.~\ref{risso}, \ref{rimsso}, and \ref{criso}. Fig.~\ref{risso} presents the contour plots of $r_{\mathrm{ISSO}}/M$ for a fixed inclination angle
$\zeta=1.25^\circ$. The smallest radius $r_{\mathrm{ISSO}}$, at which radially stable spherical orbits can exist, provides a fundamental lower bound on the allowed
location of tilted disk material and therefore plays a central role in the disk precession analysis of Sec.~\ref{sec4}.
For prograde orbits, $r_{\mathrm{ISSO}}$ decreases monotonically with the
increasing spin, reflecting the stabilizing influence of enhanced frame dragging, while it increases with magnetic field strength, indicating that the external field progressively pushes the onset of radial stability outward. In contrast, for the retrograde orbits, the trends are reversed with respect to the spin: at fixed magnetic field, $r_{\mathrm{ISSO}}$ increases monotonically as the spin grows, due to the combined effect of counter rotation and frame dragging acting against orbital stability. At fixed spin, $r_{\mathrm{ISSO}}$ also increases with the magnetic field strength, showing that magnetic effects consistently act to enlarge the inner boundary of the stable spherical motion.

In Fig.~\ref{rimsso}, we illustrate in detail the behavior of the radius $r_{\mathrm{MSSO}}$ of MSSOs in the $(a/M, BM)$ parameter space for a fixed inclination angle $\zeta = 1.25^\circ$. Owing to the fact that $r_{\mathrm{MSSO}}$ spans more than two orders of magnitude as the magnetic field strength varies, the MSSO is presented in two branches: $r_{\mathrm{MSSO1}}$, corresponding to moderate and strong magnetic fields, and $r_{\mathrm{MSSO2}}$, which captures the weak field regime approaching the Kerr limit.

For prograde orbits (left panels), the radius of the MSSOs remains relatively close to the black hole even in the presence of a strong magnetic field, reaching a minimum value of $r_{\mathrm{MSSO}} \simeq 2M$ near the upper boundary of the allowed magnetic field range. In contrast, the retrograde MSSOs (right panels) have larger radii, with a minimum value of $r_{\mathrm{MSSO}} \simeq 8M$, reflecting the combined effect of frame dragging and magnetic forces acting against orbital stability. As the magnetic field strength decreases, the magnetic effects gradually weaken and the radius of the MSSOs increases rapidly.

In the weak-field regime ($BM \lesssim 10^{-3}$), $r_{\mathrm{MSSO}}$ grows to several hundred gravitational radii, reaching $r_{\mathrm{MSSO}} \sim 290M$ at $BM \simeq 10^{-4}$. This divergence signals the recovery of the Kerr limit: in the absence of a magnetic field, the energy $E_s(r_0)$ and angular momentum $L_s(r_0)$ of spherical orbits extend monotonically to infinity, and no local maximum exists.
Consequently, the MSSOs cease to exist in the Kerr spacetime, and $r_{\mathrm{MSSO}}$ is effectively pushed to infinity.

These results demonstrate that the existence of the MSSO is a direct manifestation of the non-asymptotically flat nature of the KBR spacetime and that the magnetic field imposes a finite radial domain for spherical orbits, in sharp contrast to the Kerr case.

Finally, we present in Fig.~\ref{criso} the locus of the critical spherical orbits in the three dimensional parameter space spanned by the black hole spin $a/M$, the magnetic field strength $BM$, and the critical radius $r_{\mathrm{CSO}}/M$, which is encoded by the color scale. Each curve corresponds to the critical condition at which the swallowtail structure in the $(E_s,L_s)$ space disappears, separating the parameter regions that admit multivalued spherical orbit branches ($B < B_{\mathrm{cri}}$) from those that do not ($B > B_{\mathrm{cri}}$).

For the prograde spherical orbits shown inFig.~\ref{ELprohs}, the critical magnetic field $B_{\mathrm{cri}}$ increases monotonically with the black hole spin.
Along this curve, the color changes from deep purple to red, indicating that the corresponding critical radius $r_{\mathrm{CSO}}$ decreases from $7.5M$ for nonrotating black holes to $1.4M$ in the near extremal spin limit. This behavior shows that the high spin allows the critical spherical orbit to move deeper into the strong gravity regime, while simultaneously requiring a stronger magnetic field to eliminate the swallowtail structure. In contrast, the retrograde case shown in Fig.~\ref{ELreths} exhibits the opposite behavior. As the spin increases, the critical magnetic field decreases monotonically, whereas the color evolution along the curve indicates that the critical radius $r_{\mathrm{CSO}}$ grows from $7.5M$ to $11.3M$. This outward migration of the critical spherical orbit reflects the combined effect of frame dragging and magnetic forces acting against retrograde motion, making the disappearance of the swallowtail structure occur at larger radii.

An important consequence of this critical behavior is that, at $B=B_{\mathrm{cri}}$, the KBR spacetime admits no radially stable spherical orbits. As the blue line shown in Fig.~\ref{srEl}, both $E_s(r_0)$ and $L_s(r_0)$ become strictly monotonic functions of $r_0$, and no local minimum exists. In particular, as $r_0$ increases, the orbital angular momentum decreases and asymptotically approaches 0, in sharp contrast to the Kerr case where $L_s$ grows without bound as $r_0\to\infty$. In particular, the absence of the stable spherical orbits implies that a steady, long-lived accretion disk cannot be supported once the magnetic field exceeds the critical value. The existence of $B_{\mathrm{cri}}$ therefore imposes a stringent upper bound on the admissible magnetic field strength from the requirement of orbital stability, independently of observational or phenomenological considerations.

Therefore, these results demonstrate that the critical spherical orbit $\{a_{\mathrm{cri}}, B_{\mathrm{cri}}, r_{\mathrm{CSO}}\}$ plays a central role in organizing the structure of spherical orbits in the spacetime, providing a clear dynamical boundary between multivalued and single-valued orbital configurations.

The existence and location of these characteristic spherical orbits play a central role in shaping the structure of the tilted accretion disks, as they determine the inner boundary of the radially stable motion and directly regulate the LT precession. Using only a single observational input from M87*, namely the inclination angle $\zeta=1.25^\circ$, our analysis constrains the magnetic field parameter in the KBR spacetime to
\begin{eqnarray}
 B\lesssim 0.33 M^{-1}\quad \text{for prograde orbits},\\
 B\lesssim 0.0165  M^{-1} \quad\text{for retrograde orbits}.
\end{eqnarray}
These bounds provide essential input for the phenomenological applications discussed in the following sections.

In summary, this section establishes the dynamical framework of spherical orbits in the KBR spacetime and reveals how the presence of a uniform electromagnetic field qualitatively reshapes their stability properties. In contrast to asymptotically flat Kerr spacetime, the KBR geometry admits a finite radial domain of stable spherical orbits, characterized by the emergence of the ISSOs and MSSOs. These characteristic radii provide the fundamental dynamical scales that determine the structure of tilted accretion disks and set the stage for the LT precession analysis carried out in the following sections.

\section{Precession of spherical orbits}
\label{sec3}

Spherical orbits in rotating spacetimes are three dimensional bound motions characterized by a constant radial coordinate, polar oscillations, and secular azimuthal evolution. For generic configurations, the orbital plane is inclined by a finite angle $\zeta$ with respect to the black hole equatorial plane, resulting in a misalignment between the orbital angular momentum and the black hole spin. This misalignment induces a gravitomagnetic torque through frame dragging, leading to a secular precession of the orbital plane about the spin axis. This effect represents a natural generalization of LT precession and persists in the KBR geometry considered here.

For a spherical orbit, the radial coordinate is fixed at $r=r_0$, while the polar motion oscillates between two symmetric turning points $\theta_{\rm t}=\frac{\pi}{2}\pm\zeta$, where $\zeta$ denotes the the orbital inclination. During one complete polar oscillation between these turning points, the particle accumulates a net azimuthal advance $\Delta\varphi$ due to differential frame dragging at different latitudes. In general, this azimuthal advance does not equal $2\pi$. Consequently, the orbital plane does not close upon itself when described with respect to the Boyer-Lindquist time coordinate $t$, but instead undergoes a cumulative precession about the black hole spin axis.

Following the standard approach adopted in Refs.~\cite{Wei:2024cti, Meng:2024gcg, Meng:2025jej}, we quantify this effect by defining the generalized LT precession angular velocity as
\begin{eqnarray}
\omega_{\mathrm{LT}}
=\frac{\Delta\varphi\mp 2\pi}{T_{\theta}},
\label{eq:omegaLT}
\end{eqnarray}
where $T_{\theta}$ is the coordinate time period of the polar motion, and the upper lower sign corresponds to prograde retrograde spherical orbits. Since both $T_{\theta}$ and $\Delta\varphi$ are defined with respect to the Boyer-Lindquist time coordinate $t$, the resulting $\omega_{\mathrm{LT}}$ represents the precession frequency as measured by a distant observer.

In practice, the evaluations of $T_{\theta}$ and $\Delta\varphi$ require numerical integration of the coupled first-order geodesic equations for $\theta(\tau)$ and $\varphi(\tau)$, with the conserved energy $E_s$ and angular momentum $L_s$ fixed by Eqs.~(\ref{Es}) and (\ref{Ls}). This procedure yields the LT precession rate of the spherical orbits for arbitrary inclination angle $\zeta$, black hole spin $a$, and magnetic field parameter $B$ in the KBR spacetime. In this way, the precession frequency provides a direct dynamical link between the spherical orbit structure established in Sec.~\ref{sec2} and the observational and phenomenological implications explored in the following sections.

\section{Constraints from M87* jet precession}
\label{sec4}

In this section, we use the observed jet precession of M87 to place quantitative constraints on the black hole spin and the magnetic field strength within the KBR spacetime, based on the spherical orbit dynamics developed in Secs.~\ref{sec2} and \ref{sec3}.
The analysis proceeds by matching the observed jet precession period to the generalized LT precession of spherical orbits, and then translating the resulting parameter constraints into bounds on the radial extent of the aligned inner disk.

\subsection{Constraining KBR black hole parameters}
\label{sec4.1}

Long term VLBI monitoring of M87 has revealed quasi-periodic variations in the jet position angle, characterized by a precession period $T_{\mathrm{obs}}=11.24\pm0.47$ years, which we interpret as LT precession induced by a spinning black hole acting on a mildly misaligned accretion disk. This interpretation has been successfully employed to constrain black hole parameters in a variety of rotating spacetimes, including Kerr, Kerr-Newman, and Kerr-Taub-NUT geometries \cite{Wei:2024cti, Meng:2024gcg, Meng:2025jej}. Here, we extend this framework to the KBR spacetime, in which the electromagnetic field back-reacts nontrivially on the geometry.

We adopt a simplified but widely used warped disk model, in which the jet is launched from the vicinity of the warp radius $r_{\text{w}}$ of a thin accretion disk that precesses approximately as a rigid body under the action of frame dragging \cite{Bardeen:1975zz, Petterson:1977JA}. The disk is inclined by an angle $\zeta$ with respect to the equatorial plane, and its precession axis is assumed to coincide with the black hole spin axis. Within this picture, the jet precession frequency is identified with the generalized LT precession frequency of a spherical orbit at $r=r_{\text{w}}$, as defined in Eq.~(\ref{eq:omegaLT}). The warp radius is therefore not treated as a free parameter but is dynamically selected by the requirement of reproducing the observed precession period.

Restoring physical units, the observable precession frequency is given by
\begin{eqnarray}
\Omega_{\mathrm{LT}}= \omega_{\mathrm{LT}} \, \left(\frac{M_{\odot}}{M}\right) \left(\frac{c^3}{G M_{\odot}}\right)
\simeq 6.40982 \times 10^{12}\,
\omega_{\mathrm{LT}} \, \left(\frac{M_{\odot}}{M}\right)
\quad \mathrm{rad}\,\mathrm{yr}^{-1},
\label{eq:Omega_phys}
\end{eqnarray}
where $M_{\odot}$ denotes the solar mass and
$M = 6.5 \times 10^9 M_{\odot}$ is the mass of M87* \cite{EventHorizonTelescope:2019dse}. Throughout this section, the inclination angle is fixed at $\zeta = 1.25^\circ$, consistent with observational constraints on the jet orientation \cite{Cui:2023uyb}.

For a given set of black hole parameters $(a/M,BM)$, our procedure is as follows.
We first compute the energy $E_s(r_0,a,B,\zeta)$ and angular momentum $L_s(r_0,a,B,\zeta)$ of the spherical orbits using Eqs.~(\ref{Es}) and (\ref{Ls}). We then numerically integrate the corresponding geodesic equations to obtain $\theta(\tau)$, $\varphi(\tau)$, and $t(\tau)$, from which the LT precession frequency $\Omega_{\mathrm{LT}}$ is extracted.
By imposing the observational condition $\Omega_{\mathrm{LT}} = \frac{2\pi}{T_{\mathrm{obs}}}$, we solve for the warp radius $r_{\mathrm{w}}$ associated with each pair of parameters $(a,B)$.

Physical solutions are required to satisfy $r_{\mathrm{w}} \ge r_{\mathrm{ISSO}}$, so that the warp radius lies outside the ISSO and the disk material remains dynamically supported. In the KBR spacetime, however, the spherical orbits exhibit an additional stability property that further constrains the allowed range of $r_{\mathrm{ISSO}} \le r_{0}\mathrm{(stable)}\le r_{\mathrm{MSSO}}$. In particular, beyond the MSSO, the spherical orbits become radially unstable, in contrast to the Kerr case where stable spherical orbits extend to arbitrarily large radii. As a consequence, the physically admissible warp radius is restricted to the finite interval
\begin{equation}
r_{\mathrm{w}}\in [r_{\mathrm{ISSO}} , r_{\mathrm{MSSO}}].
\end{equation}
This upper bound arises as an intrinsic geometric effect of the KBR spacetime and provides a natural outer limit for the disk warp radius.

\begin{figure}[htbp]
\centering
\subfigure[Prograde orbits]{\label{rw_pro}
\includegraphics[width=0.45\linewidth]{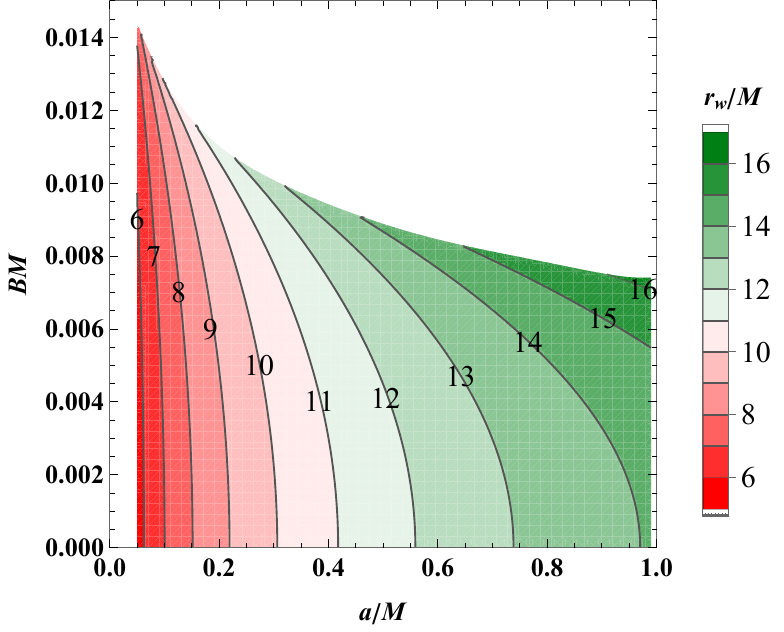}}
\subfigure[Retrograde orbits]{\label{rw_ret}
\includegraphics[width=0.45\linewidth]{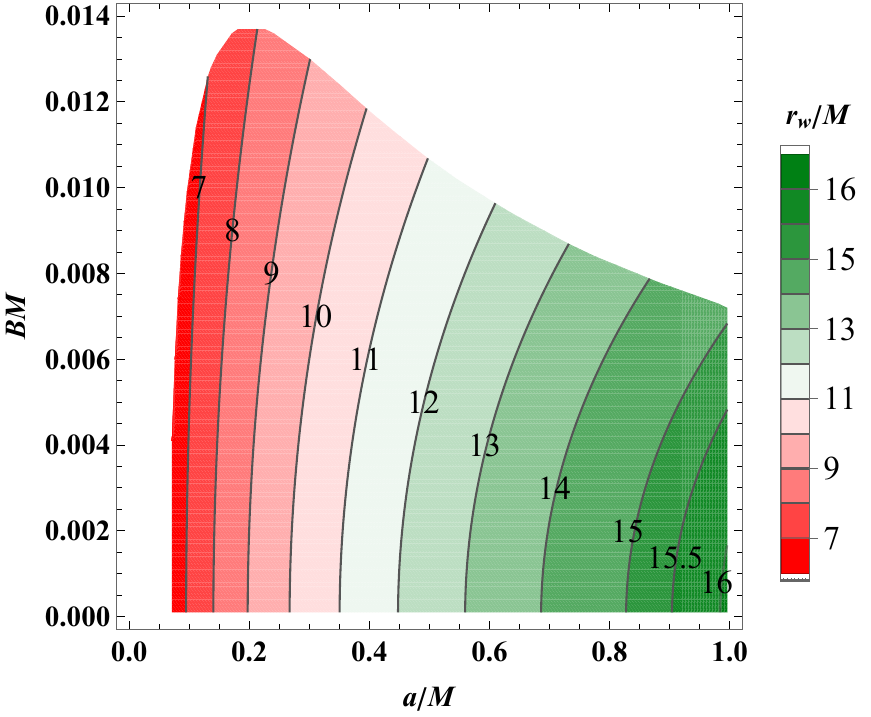}}
\caption{
Contour plots of the warp radius $r_{\text{w}}/M$ in the $(a/M, BM)$ parameter space obtained by matching the observed M87* jet precession period ($T_{\mathrm{obs}} = 11.24$ years) for spherical orbits with fixed inclination $\zeta = 1.25^\circ$.
Panels correspond to prograde and retrograde orbital orientations, respectively.
The color bar indicates the dimensionless warp radius.
Blank regions denote parameter combinations for which no stable spherical orbit exists that can simultaneously reproduce the observed precession period. Obviously,
an upper bound $BM \lesssim 0.0145$ is inferred for both orbital orientations. (a) Prograde orbital orientation. (b) Retrograde orbital orientation.
}
\label{conrw}
\end{figure}
The resulting constraint contours in the $(a/M,BM)$ parameter space are shown in Fig.~\ref{conrw} for both prograde and retrograde orientations. A key result is that the observed 11.24 years precession period imposes a robust upper bound
\begin{eqnarray}
B \lesssim 0.0145  M^{-1},
\end{eqnarray}
for both orbital configurations. For larger magnetic field strengths, the modification of the spherical orbit dynamics becomes sufficiently strong that no physically admissible solution can reproduce the observed period.

\subsection{Constraining the size of the inner disk}
\label{sec4.2}

We now translate the precession-based constraints into bounds on the geometry of the inner accretion disk. Within the standard warped disk framework \cite{Bardeen:1975zz, Wei:2024cti}, we characterize the size of the aligned inner disk by the radial separation between the warp radius $r_{\text{w}}$ and the ISCO
\begin{eqnarray}
R = r_{\text{w}} - r_{\mathrm{ISCO}}.
\label{eq:R_def}
\end{eqnarray}
Here $r_{\mathrm{ISCO}}$ is used as a conventional reference scale for the inner edge of the aligned disk, following the standard warped disk literature, while the dynamical stability of the inclined spherical orbits is ensured by the condition $r_{\text{w}} \ge r_{\mathrm{ISSO}}$ discussed in Sec.~\ref{sec4.1}.
This quantity $R$ measures the radial extent of the disk region that is both dynamically stable and approximately coplanar with the black hole equatorial plane, and is therefore directly relevant for the jet launching and near horizon accretion processes.

\begin{figure}[htbp]
\centering
\subfigure[Prograde orbits]{\label{R_pro}
\includegraphics[width=0.45\linewidth]{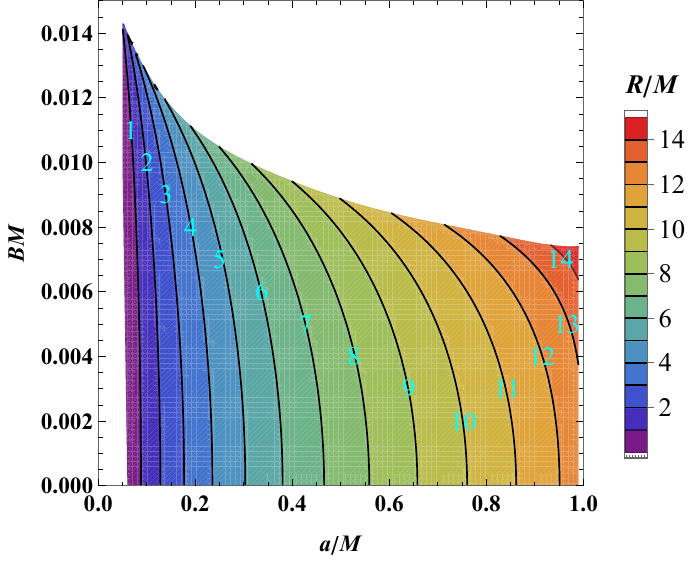}}
\subfigure[Retrograde orbits]{\label{R_ret}
\includegraphics[width=0.45\linewidth]{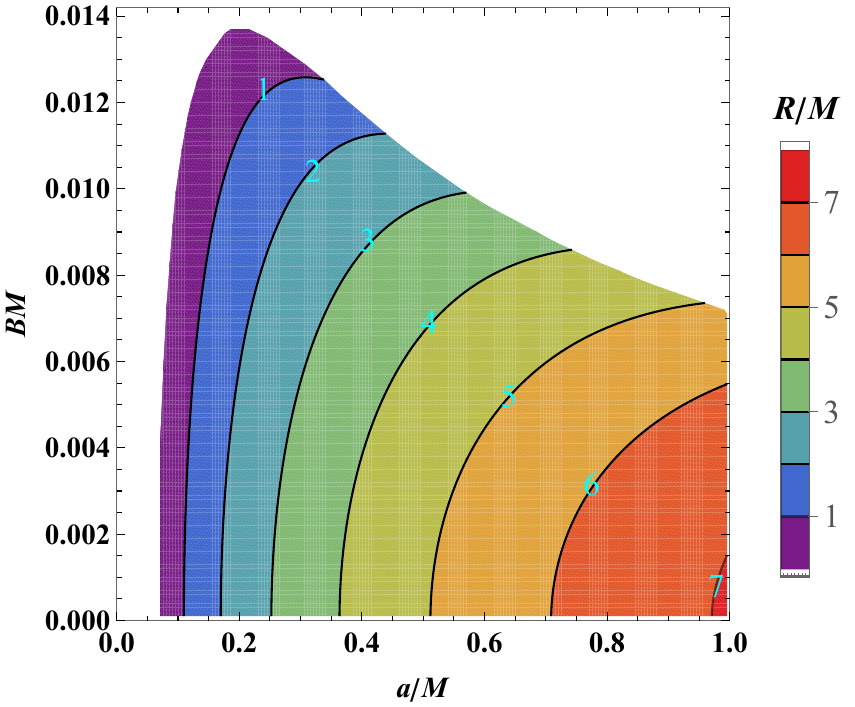}}
\caption{
Contour plots of the inner disk size $R/M = (r_{\text{w}} - r_{\mathrm{ISCO}})/M$ in the $(a/M, BM)$ parameter space, inferred from matching the M87* jet precession period ($T_{\mathrm{obs}} = 11.24$ years) for spherical orbits with fixed inclination $\zeta = 1.25^\circ$. The color bar indicates the dimensionless radial extent $R/M$.
Blank regions denote parameter combinations for which no spherical orbit solution simultaneously satisfies the precession constraint and the stability condition $r_{\text{w}} \ge r_{\mathrm{ISSO}}$. (a) Prograde orbits. (b) Retrograde orbits.
}
\label{conR}
\end{figure}

Contour plots of $R/M$ in the $(a/M,BM)$ parameter space are shown in Fig.~\ref{conR}. Several robust trends emerge. For a fixed magnetic field strength $B$, the inner disk size $R$ increases monotonically with the black hole spin $a$ for both prograde and retrograde configurations. Physically, a higher spin enhances frame dragging and increases the intrinsic LT precession rate. To reproduce the observed 11-year precession period, the warp radius must therefore shift outward to regions of weaker frame dragging. This outward displacement of $r_{\text{w}}$ dominates over the spin dependence of $r_{\mathrm{ISCO}}$, leading to a net growth of $R$ with increasing $a$.

For a fixed spin $a$, the dependence on the magnetic field strength $B$ is also monotonic in both configurations. Increasing $B$ systematically reduces the inner disk size $R$, reflecting the fact that magnetic effects modify the spherical orbit structure and precession dynamics in a way that drives the warp radius inward in order to maintain the observed precession period. This inward migration of $r_{\text{w}}$, combined with the magnetic-field-induced modification of orbital stability, leads to a progressive shrinkage of the aligned inner disk as $B$ increases.
Although the qualitative monotonic trends with respect to $a/M$ and $BM$ are shared by prograde and retrograde disks, the detailed morphology of the contour lines and the extent of the allowed parameter region differ markedly between the two cases. In particular, retrograde configurations occupy a significantly more restricted region of the $(a/M,BM)$ space, reflecting the combined influence of frame dragging, magnetic effects, and the stricter stability constraints imposed on retrograde spherical orbits.

The blank regions in Fig.~\ref{conR} correspond to combinations of $(a/M,BM)$ for which no solution satisfies both the observed precession period and the fundamental requirement $r_{\mathrm{ISSO}}\le r_{\text{w}} \le r_{\mathrm{MSSO}} $. These regions therefore represent portions of the KBR parameter space that are excluded by the M87* jet precession. Notably, the excluded regions exhibit qualitatively different structures for prograde and retrograde configurations. In certain parameter ranges, this distinction implies that the pattern of allowed and excluded regions itself can serve as a diagnostic of the sign of the disk angular momentum, offering a potential means to discriminate between prograde and retrograde accretion based purely on precession-based constraints.

By combining spherical orbit dynamics, generalized LT precession, and the observed jet precession period, we have transformed a single chronometric observable into quantitative constraints on both the black hole parameters and the structure of the inner accretion disk. Although fully self-consistent general relativistic magnetohydrodynamic simulations are ultimately required for detailed modeling, the semi-analytic framework developed here captures the dominant relativistic and magnetic effects relevant for jet precession and disk alignment, and thus provides a physically transparent and computationally efficient complement to numerical approaches.

\section{Discussions and conclusions}
\label{sec5}

In this work, we have carried out a systematic investigation of the LT precession of spherical orbits in the KBR spacetime and applied this framework to interpret the observed jet precession of M87*. Our analysis builds upon, and significantly extends, a series of recent studies that have successfully constrained black hole parameters using spherical orbit precession. The KBR spacetime introduces two essential new ingredients into this program: the absence of asymptotic flatness and the presence of a dynamically significant background magnetic field. Together, these features qualitatively alter the structure of geodesic motion and require methodological developments beyond those applicable to vacuum or asymptotically flat electrovacuum black holes.

A central technical challenge posed by the KBR geometry is the non-separability of the Hamilton-Jacobi equation for timelike geodesics, which arises from the coupled dependence of the metric functions on the radial and polar coordinates. To overcome this difficulty, we adopted a Hamiltonian formulation of geodesic motion, recasting the dynamics in terms of first-order evolution equations for both coordinates and canonical momenta. This approach provides a flexible and robust framework for tracking spherical orbits, identifying conserved quantities, and analyzing orbital stability through the conserved quantity. Within this formalism, we derived the energy $E_s$ and angular momentum $L_s$ of spherical orbits and systematically explored their dependence on the black hole spin $a$, the magnetic field parameter $B$, and the orbital inclination.

Our analysis reveals a rich orbital structure induced by the background magnetic field, most clearly manifested in the topology of the $(E_s, L_s)$ parameter space. In the Kerr limit ($B=0$), the spherical orbit family exhibits a single cusp associated with the ISSO, closely analogous to the ISCO for equatorial motion.
By contrast, in the KBR spacetime the presence of a finite magnetic field generically generates a second cusp in the $(E_s, L_s)$ space, giving rise to a characteristic swallowtail structure. This additional cusp corresponds to a MSSO, which has no counterpart in the asymptotically flat Kerr geometry and therefore represents a distinctive dynamical feature of the KBR spacetime.

The two cusps delimit a finite radial interval within which spherical orbits are radially stable, while their merger at a critical magnetic field $B_{\mathrm{cri}}$ defines a critical spherical orbit. This sequence of magnetic field driven topological transitions provides a compact geometric encoding of orbital stability in the KBR spacetime and has no analogue in asymptotically flat black hole geometries.
Importantly, when $B \ge B_{\mathrm{cri}}$, the swallowtail structure disappears and the functions $E_s(r_0)$ and $L_s(r_0)$ become monotonic. As shown explicitly in Fig.~\ref{srEl}, although spherical orbits still exist in this regime, they are radially unstable at all radii and therefore cannot support a stable accretion disk configuration.

As a direct consequence, the existence of a dynamically significant magnetic field in the KBR geometry imposes a strict upper bound on the allowed magnetic field strength for the stable disk formation. The critical magnetic field $B_{\mathrm{cri}}$, computed in Fig.~\ref{criso}, thus acquires a clear physical interpretation: it marks the threshold beyond which stable spherical orbits and hence stable tilted accretion disks-cease to exist. In this sense, the magnetic field in the KBR spacetime plays a dual role, simultaneously reshaping the orbital structure and tightly constraining the conditions under which accretion disks can remain stable.

Building on these orbital properties, we constructed a phenomenological model of the tilted accretion disk in the M87*, in which the disk particles are assumed to occupy spherical orbits with an inclination fixed to the observed jet misalignment angle, $\zeta = 1.25^\circ$ \cite{Cui:2023uyb}. This observational input is consistently imposed throughout our analysis and is used in determining the characteristic spherical orbits of the KBR spacetime, including the ISSO, the MSSO, the CSO, and the warp radius $r_{\mathrm{w}}$. Already at this stage, the mere requirement that spherical orbits with $\zeta = 1.25^\circ$ exist and remain well defined imposes a first, nontrivial constraint on the black hole parameters, restricting the magnetic field strength to $BM \lesssim \mathcal{O}(10^{-1})$ for prograde motion and $BM \lesssim \mathcal{O}(10^{-2})$ for retrograde motion.

A further constraint on the warp radius arises from the radial stability properties of spherical orbits in the KBR spacetime.
Because spherical orbits are radially stable only within the finite interval between the ISSO and the MSSO, the physically admissible warp radius is required to satisfy
$r_{\mathrm{ISSO}} \le r_{\mathrm{w}} \le r_{\mathrm{MSSO}}$.
This condition provides a second, purely geometric restriction on $r_{\mathrm{w}}$, which relies on the observed inclination angle but does not yet invoke any timing information and therefore does not further narrow the black hole parameter space.

Finally, by identifying the observed jet precession period, $T_{\mathrm{obs}} = 11.24$ years, with the generalized LT precession period evaluated at $r_{\mathrm{w}}$, we obtain a second and more stringent constraint on both the warp radius and the KBR black hole parameters. Remarkably, this precession based analysis sharpens the allowed parameter space into a quantitative upper bound on the magnetic field strength, $B \lesssim 0.0145\,M^{-1}$, which applies to both prograde and retrograde configurations. Notably, this bound depends only on the jet precession period and the disk inclination, and is therefore largely insensitive to the detailed microphysics of the accretion process. Its consistency with independent constraints derived from black hole shadow observations \cite{Wang:2025vsx, Zeng:2025tji, Ali:2025beh, Vachher:2025jsq} provides complementary support for the presence of a moderately strong, large-scale magnetic field in the immediate environment of M87*.

Beyond parameter constraints, we also characterized the geometry of the inner accretion disk by evaluating the radial extent of the aligned region, defined as $R = r_{\text{w}}-r_{\mathrm{ISCO}}$. The resulting contours in the $(a/M,BM)$ parameter space exhibit clear and robust trends. For a fixed magnetic field strength, $R$ increases monotonically with the black hole spin for both prograde and retrograde configurations, reflecting the outward shift of the warp radius required to reproduce the observed precession period in the presence of stronger frame dragging.
For a fixed spin, increasing the magnetic field strength generally reduces $R$, as magnetic effects drive the warp radius inward and restrict the domain of stable spherical orbits. While these monotonic dependencies on $a$ and $B$ are common to both disk orientations, the allowed parameter regions differ substantially between prograde and retrograde cases, with retrograde configurations occupying a more restricted portion of the $(a,B)$ plane.
These qualitative differences suggest that observational constraints on disk orientation, when combined with precession measurements, could provide valuable additional leverage for discriminating between viable regions of parameter space in magnetized black hole systems.

In summary, this work demonstrates the effectiveness of jet precession as a precision probe of strong field gravity in magnetized black hole spacetimes. By unifying spherical orbit dynamics, Hamiltonian geodesic analysis, and observational input from M87*, we have:
\begin{enumerate}
    \item established a practical theoretical framework for treating non-separable geodesic motion in complex, magnetized spacetimes;
    \item identified distinctive topological signatures in orbital parameter space that are intrinsic to non-asymptotically flat black hole geometries;
    \item derived new, independent constraints on the magnetic field strength surrounding M87* through a two-stage, observation-driven analysis;
    \item extended the spherical orbit precession method into a regime where external magnetic fields play a dynamically essential role.
\end{enumerate}

These results highlight the power of combining relativistic orbital dynamics with high precision astrophysical timing observations. Future measurements of jet precession in M87* and other active galactic nuclei, together with advances in general relativistic magnetohydrodynamic simulations and multi-wavelength observations, will enable increasingly stringent tests of magnetized black hole spacetimes and deepen our understanding of the interplay between black holes, accretion flows, and magnetic fields in the strong gravity regime.

\section*{Acknowledgements}
This work was supported by the National Natural Science Foundation of China (Grants No. 12475055, and No. 12247101), the Fundamental Research Funds for the Central Universities (Grant No. lzujbky-2025-jdzx07),
and the Natural Science Foundation of Gansu Province
(No. 22JR5RA389, No.25JRRA799).

\end{document}